\documentclass[sigconf, nonacm]{acmart}

\usepackage{amsmath,amssymb,amsfonts}
\usepackage{graphicx}
\usepackage{textcomp}
\usepackage{xcolor}
\usepackage{listings}
\usepackage{color}
\usepackage{hyperref}

\definecolor{codegreen}{rgb}{0,0.6,0}
\definecolor{codegray}{rgb}{0.5,0.5,0.5}
\definecolor{codepurple}{HTML}{C42043}
\definecolor{backcolour}{HTML}{F2F2F2}
\definecolor{bookColor}{cmyk}{0,0,0,0.90}  

\usepackage{listings}
\usepackage{xcolor}
\usepackage{algorithm}
\usepackage{algpseudocode}
\usepackage{amsmath}
\usepackage{bm}
\usepackage{multirow}
\usepackage{float}
\usepackage[section]{placeins}
\usepackage{subcaption}
\usepackage[normalem]{ulem}
\usepackage{balance}
\usepackage{tabularx}

\def\BibTeX{{\rm B\kern-.05em{\sc i\kern-.025em b}\kern-.08em
    T\kern-.1667em\lower.7ex\hbox{E}\kern-.125emX}}

\newcommand\vldbdoi{XX.XX/XXX.XX}
\newcommand\vldbpages{XXX-XXX}
\newcommand\vldbvolume{14}
\newcommand\vldbissue{1}
\newcommand\vldbyear{2020}
\newcommand\vldbauthors{\authors}
\newcommand\vldbtitle{\shorttitle} 
\newcommand\vldbavailabilityurl{https://github.com/rebornDanny/CoeusBI}
\newcommand\vldbpagestyle{plain} 

\begin{document}

\title{\vspace{-0.02cm}CoeusBI: A Comprehensive Interactive Business Intelligence System Powered by LLMs at Baidu [Extended Version]}

\author{
  Jinqing Lian{$^{\dagger\star}$},~~~Chaofan Li{$^{\dagger}$},~~~Yingxia Shao{$^{\dagger*}$},~~~Ming Wang{$^{\ddagger}$},~~~Yang Dong{$^{\ddagger}$},~~~Xinyi Liu{$^{\ddagger}$},~~~Wei Zhang{$^{\ddagger}$},~~~Chaoxian Gui{$^{\ddagger}$},~~~Tianqi Wan{$^{\ddagger\star}$},~~~Ming Dong{$^{\ddagger*}$}
}
\affiliation{%
  \institution{
  {$^{\dagger}$} Beijing University of Posts and Telecommunications \\
  {$^{\ddagger}$} Baidu Inc. \\        
  }
}
\email{{jinqinglian,cfli,shaoyx}@bupt.edu.cn, {wangming15,dongyang06,liuxinyi02,zhangwei143,guichaoxian,dongming03}@baidu.com} 
\thanks{This work was done while $^\star$Jinqing Lian and $^\star$Tianqi Wan worked at Baidu.}
\thanks{$^*$Yingxia Shao and $^*$Ming Dong are the corresponding authors.}

\begin{abstract}
The advent of Large Language Models (LLMs) has catalyzed the emergence of interactive Business Intelligence (BI) systems. Although commercial BI products increasingly adopt semantic layers paired with natural language interfaces, they predominantly rely on manual configurations to define metrics and dimensions. Real-world deployments continue to face critical challenges: (a) frequent JOIN operations degrade the accuracy of SQL generation; (b) wide schemas exacerbate the challenge of schema linking; and (c) the generation of dialect-specific queries and the accurate support for multi-round dialogues incur high computational costs and yield limited accuracy. 

We introduce CoeusBI, an industrial-scale interactive BI system that addresses these barriers through a novel Dual-Agent Architecture paired with a Hierarchical Schema Linking module: (1) an offline View Generation Agent that utilizes error-feedback to autonomously convert complex JOIN queries into simple single-view queries, which eliminates the need for manual semantic modeling; (2) a Hierarchical Schema Linking module that leverages vector retrieval over views to handle exceptionally wide schemas efficiently; and (3) a dynamic Routing Agent that evaluates dialogue contexts to route queries, dynamically invoking either the synthesis of new intermediate representations or targeted modifications of existing ones, before compiling the unified representation via a deterministic SQL compiler that is agnostic to dialects. 

Extensive experiments on both public datasets and production datasets demonstrate that CoeusBI achieves significant improvements in query accuracy, token efficiency, and user satisfaction relative to existing methods. CoeusBI is currently deployed as a standalone service on the data platform of Baidu and is widely used across multiple business lines—including video, search, and advertising—supporting thousands of users daily, thereby evidencing strong practicality and scalability. 
\end{abstract}

\maketitle

\color{bookColor}

\pagestyle{\vldbpagestyle}
\begingroup\small\noindent\raggedright\textbf{PVLDB Reference Format:}\\
\vldbauthors. \vldbtitle. PVLDB, \vldbvolume(\vldbissue): \vldbpages, \vldbyear.\\
\href{https://doi.org/\vldbdoi}{doi:\vldbdoi}
\endgroup
\begingroup
\renewcommand\thefootnote{}\footnote{\noindent
This work is licensed under the Creative Commons BY-NC-ND 4.0 International License. Visit \url{https://creativecommons.org/licenses/by-nc-nd/4.0/} to view a copy of this license. For any use beyond those covered by this license, obtain permission by emailing \href{mailto:info@vldb.org}{info@vldb.org}. Copyright is held by the owner/author(s). Publication rights licensed to the VLDB Endowment. \\
\raggedright Proceedings of the VLDB Endowment, Vol. \vldbvolume, No. \vldbissue\ %
ISSN 2150-8097. \\
\href{https://doi.org/\vldbdoi}{doi:\vldbdoi} \\
}\addtocounter{footnote}{-1}\endgroup

\ifdefempty{\vldbavailabilityurl}{}{
\vspace{.3cm}
\begingroup\small\noindent\raggedright\textbf{PVLDB Artifact Availability:}\\
The source code, data, and/or other artifacts have been made available at \url{\vldbavailabilityurl}.
\endgroup
}

\section{Introduction}

Business Intelligence (BI) systems~\cite{becker2019microsoft, carlisle2018software, chaudhuri2011overview, linkon2024advancements, nuseir2021designing, hegger2024smart, sukhdeve2023google, negash2008business, wieder2015impact} are widely deployed at Baidu and comparable companies to analyze raw data generated by diverse business lines. 
They span the entire lifecycle, from data storage, organization, and modeling to analysis and presentation. 
Through the systematic analysis of raw data and the clear dissemination of results, BI systems provide robust support for data-driven decision-making. 

In recent years, artificial intelligence has advanced rapidly, with Large Language Models (LLMs) achieving notable gains in natural language understanding. 
Building on these advances, interactive BI systems leverage LLMs to translate user queries expressed in natural language into corresponding analytical results. 
State-of-the-art commercial systems, such as Snowflake Cortex Analyst~\cite{anoshin2025snowflake}, Databricks Genie~\cite{gupta2024databricks}, and Google BigQuery~\cite{lakshmanan2019google} with Gemini, increasingly employ a semantic layer to bridge the gap between natural language and database schemas. 
However, these commercial solutions predominantly depend on manual configurations (e.g., YAML definitions or LookML) to define metrics, dimensions, and relationships. 
This manual modeling incurs substantial human labor and limits adaptability when schemas evolve.

\begin{table}[t]
\begin{center}
\caption{Table and column statistics of public datasets and production datasets.}
\label{table1}
\begin{tabular}{lccc}
\noalign{\hrule height 1.2pt}
Dataset                & \multicolumn{1}{l}{Tables} & \multicolumn{1}{l}{Columns} & \multicolumn{1}{l}{Avg. Columns} \\ \hline
\multicolumn{4}{c}{NL2SQL Public Dataset}                                                                          \\ \hline
WikiSQL                & 26521                      & $\sim$ 125000               & 4.7                              \\
Spider                 & 1020                       & $\sim$ 5520                 & 5.1                              \\
BIRD                   & 694                        & $\sim$ 7600                 & \textbf{10.95}                    \\ \hline
\multicolumn{4}{c}{NL2BI Production Dataset}                                                                       \\ \hline
MS-Financial           & 632                        & 4000                        & 6.3                              \\
MS-Financial-Views     & 200                        & 7400                        & 37                               \\
MS-Commercial          & 2281                       & 65000                       & 28.5                             \\
MS-Commercial-Views    & 1600                       & 50000                       & 31.3                             \\
BD-Baijiahao-Views     & 52                         & 11388                       & \textbf{219}                     \\
BD-Baidu App-Views     & 20                         & 3060                        & 153                              \\
BD-Search-Views        & 43                         & 7792                        & 181.2                            \\
BD-Haokan Video-Views & 33                         & 3018                        & 91.5                             \\ 
\noalign{\hrule height 1.2pt}
\end{tabular}
\end{center}
\vspace{-2.5em}
\end{table}

Furthermore, the deployment of an interactive BI system in a real-world production environment encounters several fundamental challenges that distinguish it from standard academic benchmarks~\cite{leispider, pourreza2024din, wang2025mac, li2024can, yu2018spider, gao2024xiyan, liu2024survey, WikiSQL, shkapenyuk2025automatic, lee2025mcs, zhou2024db, liu2025xiyan, dharalarge}.

\noindent
\textbf{C1: Frequent JOIN operations degrade the accuracy of SQL generation and the efficiency of query execution.} 
In practical BI environments, a substantial proportion of queries involve JOIN clauses over numerous base tables. 
We observe that 90.8\% of golden SQL queries contain JOIN operations when we apply the classifier of DIN-SQL~\cite{pourreza2024din} to the dataset of BD-Business Line A at Baidu (As shown in Table~\ref{tableview}). 
The generation of SQL queries with multi-table JOIN clauses is substantially more difficult and limits accuracy~\cite{nascimento2026text}. 

\noindent
\textbf{C2: Wide schemas exacerbate the challenge of schema linking.}
As illustrated in Table~\ref{table1}, industrial datasets possess significantly wider schemas than public datasets. 
The introduction of views further increases the width of the schemas, where the BD-Baijiahao dataset averages 219 columns per view. 
Prompt-based schema linking methods~\cite{gao2024xiyan, liu2025xiyan} fail to scale in such environments, quickly exhausting token budgets. 

\noindent
\textbf{C3: The accurate handling of multi-round dialogues and the generation of dialect-specific queries incur high costs and exhibit limited predictability.} 
Interactive BI systems process two primary categories of user interactions: Single-Round Dialogues (SRD), which represent self-contained questions requiring a complete schema-linking and query synthesis pass, and Multi-Round Dialogues (MRD), which represent follow-up questions that refine or extend the previous turn. 
Production settings demand support for multiple dialects (e.g., ClickHouse, MySQL), which introduces non-trivial costs for training or inference~\cite{pourreza2024dts, wang2025mac}. 
Concurrently, prevailing methods (such as SiriusBI~\cite{jiang2024siriusbi}) employ query rewriting to manage MRDs and rely on LLMs to generate the final SQL queries, creating a risk of hallucination and reducing overall predictability.

To address the aforementioned challenges, we introduce CoeusBI, an interactive BI system powered by LLMs deployed in the production environment of Baidu. 
Unlike commercial solutions that require manual configurations, CoeusBI automates the creation of semantic views to convert complex JOIN operations into simple queries via the View Generation Agent. 
It incorporates a Hierarchical Schema Linking module tailored for wide schemas and implements a Routing Agent paired with a deterministic SQL compiler to ensure accurate support for multi-round dialogues and eliminate hallucination risks in the final SQL generation step. 
CoeusBI serves thousands of users within Baidu, and the strong performance on both public datasets and production datasets demonstrates significant practical effectiveness and scalability. 

Specifically, we make the following contributions:
\begin{itemize}
\item {} We introduce a \textbf{Dual-Agent Architecture} comprising an autonomous \textbf{View Generation Agent} that transforms a wide range of queries that include JOIN operations into simple single-view queries, eliminating the dependency on manual semantic configurations. 
\item {} We introduce a \textbf{Hierarchical Schema Linking module} to mitigate the challenge of wide schemas, yielding high retrieval accuracy while minimizing token overhead. 
\item {} We present a \textbf{Routing Agent} and a \textbf{deterministic SQL compiler} that ensure support that is agnostic to dialects, eliminate the hallucination risks inherent in SQL generation based on LLMs, and dynamically route conversational queries. 
\item {} CoeusBI achieves high accuracy and low operational costs on both public datasets and the production datasets of Baidu. 
\end{itemize}

\section{Preliminary}

\begin{figure}[htbp]
  \includegraphics[width=0.485\textwidth]{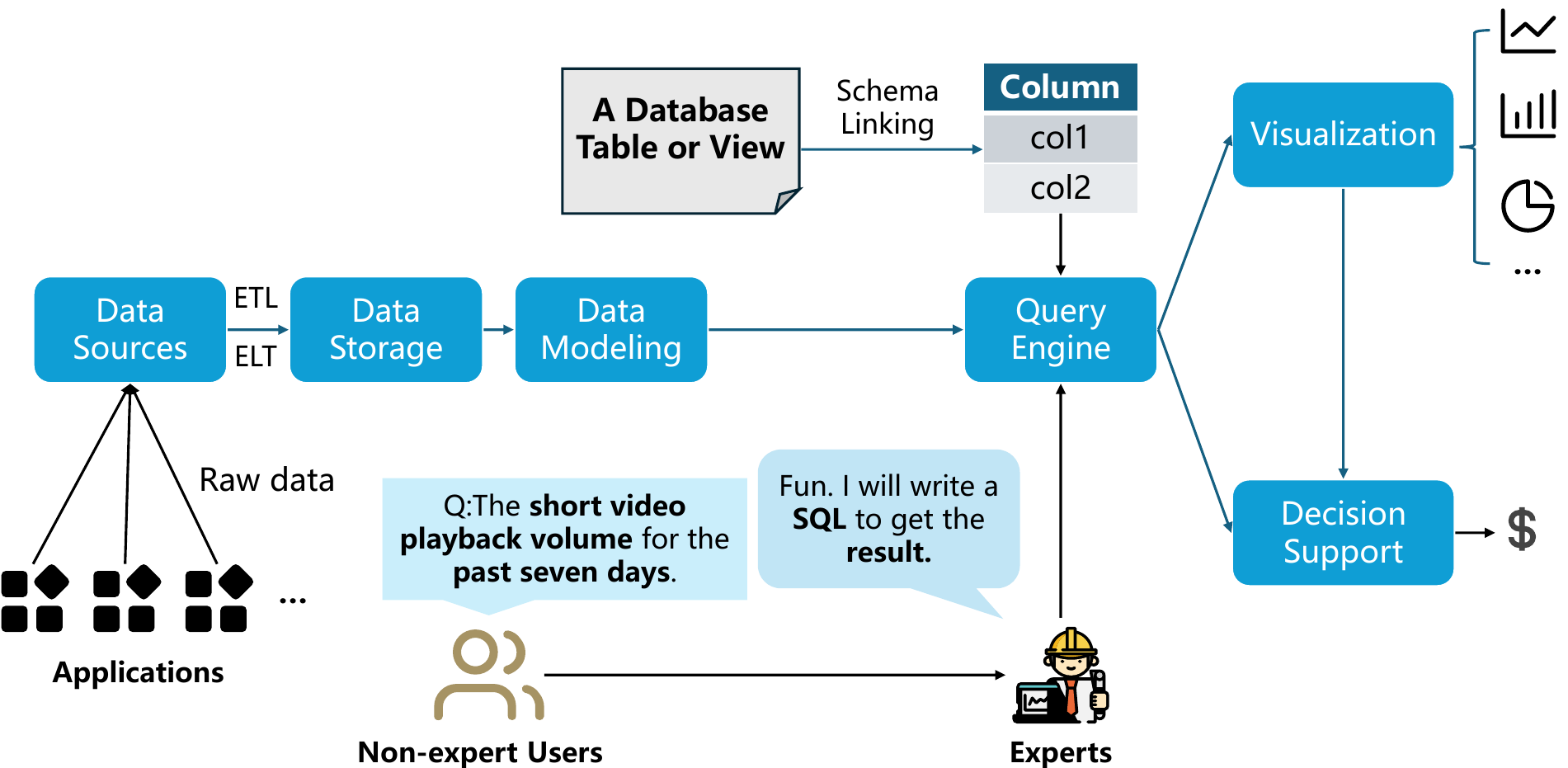}
  \caption{Example BI Pipeline. }
  \label{figure:exampleOfBI}
\end{figure}

\begin{figure*}[htbp]
  \includegraphics[width=0.90\textwidth]{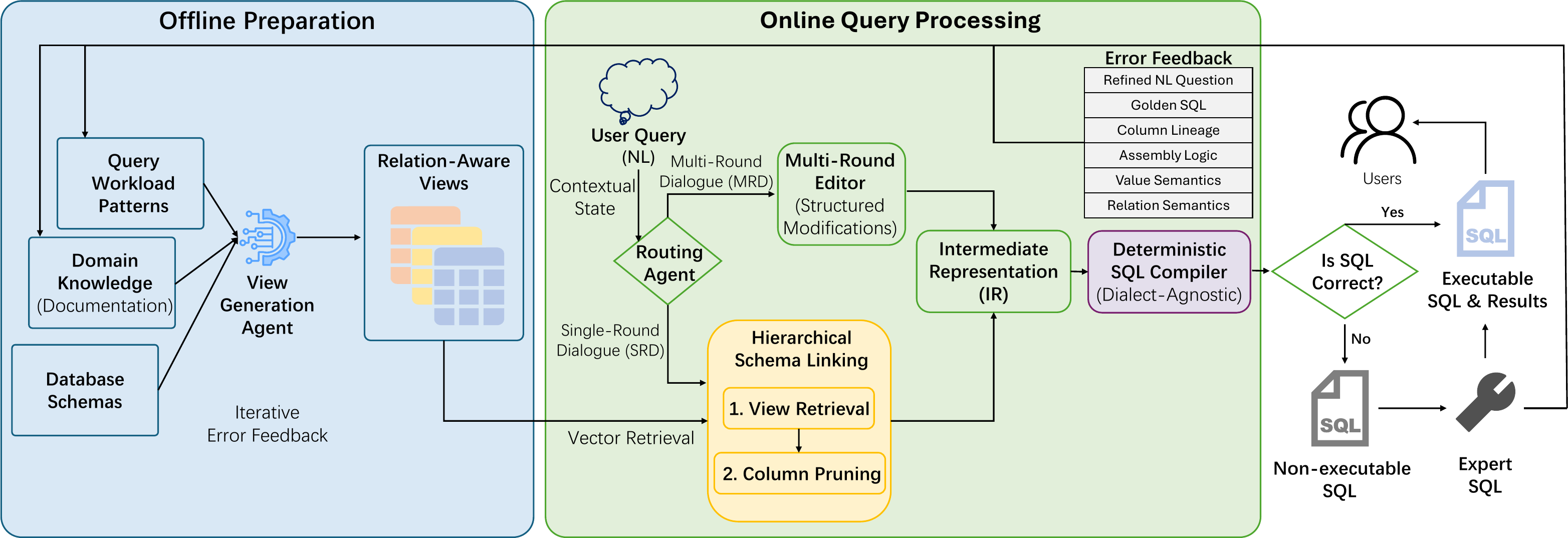}
  \caption{Overview of CoeusBI.}
  \label{figure:overviewOfCoeusBI}
  \vspace{-1em}
\end{figure*}

\subsection{Business Intelligence Background}
As illustrated in Figure~\ref{figure:exampleOfBI}, a typical BI system consists of six core components.
\textbf{Data Sources Layer}: This layer gathers raw data from various sources. 
\textbf{Data Storage Layer}: Data from the source layer is ingested via Extract-Transform-Load processes and stored in systems such as data warehouses. 
\textbf{Data Modeling Layer}: This component organizes and models the ingested data to facilitate downstream query and analysis tasks.
\textbf{Query Engine Layer}: This layer handles user queries, aggregations, and computations over the data. 
\textbf{Visualization Layer}: Processed query results are presented through dashboards and reports. 
\textbf{Decision Support Layer}: The final insights are delivered to end users to support business decision-making.

\subsection{Interactive Business Intelligence Systems}
Interactive BI systems provide analysts with an analytics experience centered on iterative refinement and conversational interaction. 
These systems translate natural language into precise analytical operations. 
Interactive BI systems emphasize continuous collaboration with business users, who expect accurate BI outcomes under constrained input budgets and complex data schemas.

\subsection{Schema Linking}
Schema linking is a critical step in NL2SQL and NL2BI that prunes wide schemas down to the essential tables and columns required for a specific query, thereby dramatically improving the accuracy and efficiency of SQL generation. 
In the era of LLMs, where limitations of the context window restrict the input length, the role of schema linking has become even more pivotal.

\section{Overview of CoeusBI}\label{sec:overview}

Figure~\ref{figure:overviewOfCoeusBI} illustrates the system architecture of CoeusBI, which is distinctly divided into two main environments: \textbf{Offline Preparation} and \textbf{Online Query Processing}.

In the \textbf{Offline Preparation} phase, domain knowledge (sourced from documentation) and raw database schemas are ingested by the \textbf{View Generation Agent}. Employing an iterative error-feedback mechanism, this agent automatically processes these inputs to construct \textbf{Relation-Aware Views}. This offline step effectively handles the intricacies of complex data relationships beforehand, structuring the views to be readily accessible via vector retrieval during the online execution.

During \textbf{Online Query Processing}, the workflow is initiated by a natural language (NL) user query. A central \textbf{Routing Agent} evaluates the input to dynamically direct the execution path into either an SRD branch or an MRD branch:
\begin{itemize}
\item \textbf{New Query (SRD):} The request represents a self-contained intent and is routed to the \textbf{Hierarchical Schema Linking} module. This module interacts directly with the offline-generated Relation-Aware Views via vector retrieval, executing a precise two-step process: (1) View Retrieval, followed by (2) Column Pruning.
\item \textbf{Multi-Round Dialogue (MRD):} If conversational context is detected, indicating a refinement of the preceding interaction, the query is routed to the \textbf{Multi-Round Editor}. This module leverages the contextual state from previous intermediate representation to apply structured modifications, efficiently bypassing the full schema linking process.
\end{itemize}

Crucially, both routing paths converge to generate a unified \textbf{Intermediate Representation (IR)}. This IR is then passed to a \textbf{deterministic SQL compiler}. Because this compiler operates in a dialect-agnostic manner, it reliably translates the IR into \textbf{Executable SQL \& Results} while explicitly eliminating any hallucination risks associated with SQL generation based on LLMs. To handle unresolved intents or compilation errors seamlessly, the system avoids unconstrained language models, routing these edge cases to an expert interface. Experts author correct executable queries, accumulated by the system as feedback artifacts. Maintaining stability, the system incorporates these artifacts into an offline error feedback cycle, batch-updating views when error frequencies exceed a predefined threshold. Coupled with atomic schema deployments, this batching strategy preserves pipeline determinism while continuously refining the system. Analytical results from expert queries naturally inform subsequent multi-round interactions.

\section{View Generation Agent}\label{Section:section4}

Recent work has shown that lightweight semantic descriptions can aid NL2SQL systems~\cite{eben2025rasl, XiyanDBDescGen}. 
However, the use of per-query auxiliary descriptions at the Query Engine Layer imposes substantial computational overhead. 
We generate relation-aware descriptions to automatically construct views that convert JOIN-intensive BI queries into single-view queries via an autonomous agentic workflow. 
This automated approach differentiates CoeusBI from commercial solutions that require manual semantic view definitions. 
Our agent comprises two components: a description generator that produces semantic summaries capturing query patterns, and a view generator that translates these descriptions into executable views.

\subsection{Description Generator}\label{sec4:descriptionGenerator}

Domain knowledge encompasses jargon (e.g., `DAU' for daily active users, `DNU' for daily new users, `LTV' for lifetime value), codified calculation conventions (e.g., employing month-to-date, MTD, when computing period-over-period metrics in certain business lines), and default query settings (e.g., implicitly including the data of the current day). 
Because such knowledge is routinely reflected in user queries, interactive BI systems must accurately model and interpret it to ensure accurate responses. 

The description generator incorporates domain knowledge—crucial in BI contexts—to enrich the generation of the descriptions. 
Concretely, it takes both the database schema and domain knowledge as inputs. 
Specifically, the system employs a static prompt template directing the LLM to extract the tables' main purpose, typical query patterns, and key relationships. The description generator autonomously infers the necessary content to satisfy these extraction goals. Without domain knowledge, the system avoids upfront manual construction. Instead, the LLM relies on linguistic clues, such as table and column names, alongside structural constraints within the raw schema to infer query purposes and JOIN paths between tables. Conversely, available domain knowledge derives from three primary sources: existing internal documentation detailing business jargon, per-table descriptive files accompanying public datasets, and error feedback artifacts continuously accumulated by experts during online failures. This supplementary domain knowledge enables the generator to capture proprietary terminology and latent relations among business-specific aggregation dimensions.
Within the BIRD-dev dataset, we employ the descriptions of the dataset per table—for example, the contents of event.csv for table `event'—as domain knowledge for the synthesis of the descriptions.
However, the breadth of this domain knowledge is substantially more limited than the domain knowledge found in real-world BI settings.
For datasets without domain knowledge (e.g., Spider), the domain knowledge in the prompt is left empty. 

The prompt is as follows:

\begin{lstlisting} [language = c++]
/* You are a database expert creating semantic table descriptions for a text-to-SQL and text-to-BI retrieval system.  
### TASK 
Generate each concise, high-level description of the provided database tables that captures its semantic purpose and usage patterns. This description will be embedded in a prompt and used for retrieving relevant tables when processing natural language questions. 
### DATABASE SCHEMA {} 
### DOMAIN KNOWLEDGE {} 
First, think through what makes these tables important, what business concepts they represent, and how users might refer to them in natural language questions. Consider its likely role in the database without listing all columns. Then, create each concise table description that covers: 
1. The table's main purpose and real-world concept it represents 
2. Typical query patterns or business questions it helps answer 
3. Key relationships with other tables (if any) 
Keep your description under 300 words, focusing on semantic meaning rather than technical details. For tables with many columns, focus on the overall table purpose and categories of data rather than describing individual columns. 
Format your response as follows: <description>.  */
\end{lstlisting}

The generated example output is illustrated below:

\begin{lstlisting} [language = c++]
/* Your analysis of the table and reasoning about its purpose, usage, and significance.
This table records organizational events, capturing what the event is, when and where it occurs, its type, and current lifecycle status. It is a central planning and reporting entity for activities such as meetings, games, socials, or elections.
Users ask for upcoming or past events, events by type or location, and status-based lists (planning/open/closed). It supports questions like "What events happened last month?" or "Where is the next game?" and "Show notes for the election."
It connects to attendance (who participated) and budget/expense (financials per event) via budget lines. */
\end{lstlisting}

This example shows a generated description for the `event' table in the student\_club database of the BIRD-dev dataset. 
The description captures both user intent (``Users ask for upcoming or past events, events by type or location, and status-based lists (planning/open/closed).'') and the JOIN paths of the table (``It connects to attendance (who participated) and budget/expense (financials per event) via budget lines.''). 
In these JOIN paths, `Attendance', `Budget', and `Expense' are table names, whereas `budget lines' denotes a column within a table.
Within the BIRD-dev dataset, we observe that even when the description generator leverages limited domain knowledge alongside the complete database schema, the semantic descriptions remain highly interpretable for understanding queries that include JOINs within the database.

\subsection{View Generator}

By leveraging domain knowledge and predicting the JOIN strategies employed by natural language queries (e.g., ``What events happened last month?'') over the underlying base tables, the description generator ensures that semantic descriptions contain valid JOIN paths. The view generator subsequently treats these semantic descriptions as database-level specifications to generate views. This process yields views that naturally encode JOIN semantics, thereby effectively accommodating queries with JOIN clauses. Supporting multiple SQL dialects requires only an explicit dialect specification within the prompt. For instance, when targeting the SQLite dialect used in the BIRD-dev dataset, the prompt for view generation is formulated as follows:

\begin{lstlisting} [language = c++]
/* You are a database expert creating SQLite views for a text-to-SQL and text-to-BI retrieval system. 
### TASK 
Generate SQLite views of the provided database table descriptions and tables. 
### DATABASE SCHEMA {} 
### DATABASE DESCRIPTION {} 
First, think through how users might refer to it in natural language questions. It is expected that subsequent natural language queries can be performed solely through your generated view without accessing the original tables, meaning all queries should be able to obtain results directly via your generated view. Using DATABASE SCHEMA and DATABASE DESCRIPTION to generate the SQLite views. 
Format your response as follows: 
The SQLite view: <SQL>.  */
\end{lstlisting}

Crucially, the view generation process is demand-driven rather than structure-driven, which actively prevents a combinatorial explosion of possible views in large-scale enterprise data warehouses. By leveraging domain knowledge and acknowledging the empirical locality of enterprise queries—which heavily cluster around specific business themes—the View Generation Agent predicts prevalent thematic JOIN paths instead of enumerating arbitrary structural combinations. Consequently, the total number of generated views is naturally bounded by the diversity of query demands, enabling the agent to consolidate multiple base tables into coherent, subject-oriented views while excluding uninvolved tables.

To capture unpredictable cross-domain requirements, the View Generation Agent leverages the batched error feedback mechanism introduced in Section~\ref{sec:overview}. Rather than regenerating views upon every single error, the agent utilizes the accumulated expert-authored artifacts to refine view definitions during the scheduled offline updates, thereby preventing frequent cache invalidation of the schema linking module.

We generate 84 views across the 11 databases in the BIRD-dev dataset. The generation of views for the BIRD-dev dataset required an offline processing time of only 22.5 minutes using GPT-5, demonstrating that the cost of generation is exceedingly low. All views are publicly available at \href{https://github.com/rebornDanny/CoeusBI}{githubRepo}. 
To ensure experimental fairness, the iterative refinement mechanism driven by execution errors is disabled during all evaluations and during the generation of views for the BIRD-dev dataset.

Table~\ref{tableview} presents the query categorization on the BIRD-dev dataset using the classifier of DIN-SQL before and after view generation. The View Generation Agent \textbf{converts a wide range of JOIN-intensive queries into single-view queries}. Prior to the introduction of views, 77.03\% of the queries contain JOIN clauses. Following view generation, this proportion decreases to \textbf{47.61\%}. The share of \textit{Easy} queries increases from 22.97\% to \textbf{52.39\%}, and the overall execution accuracy of DIN-SQL + DeepSeek-R1-0528 improves from 61.67\% to \textbf{63.40\%}. We perform the same analysis on a production dataset, BD-Business Line A, where the proportion of queries including JOIN clauses declines from 90.8\% to \textbf{25.72\%}, and the execution accuracy of DIN-SQL + DeepSeek-R1-0528 increases from 6.03\% to \textbf{23.49\%}.

\begin{table}[t]
\begin{center}
\caption{Comparative distribution of query categories and the corresponding execution accuracy of DIN-SQL with and without the auto-generated views from the View Generation Agent of CoeusBI on two datasets (The execution of DIN-SQL utilizes DeepSeek-R1-0528 as the underlying model; Easy queries do not contain JOINs; Non-nested and Nested queries include JOINs; EX = execution accuracy).}
\label{tableview}
\begin{tabular}{lcccc}
\noalign{\hrule height 1.2pt}
Method                        & Easy             & Non-nested       & Nested           & EX               \\ \hline
\multicolumn{5}{c}{BIRD-dev}                                                                            \\ \hline
DIN-SQL        & 22.97\%          & 64.55\%          & 12.48\%          & 61.67\%          \\
 \quad w/ views & \textbf{52.39\%} & \textbf{35.83\%} & \textbf{11.78\%} & \textbf{63.40\%} \\ \hline
\multicolumn{5}{c}{BD-Business Line A}                                                                  \\ \hline
DIN-SQL        & 9.21\%           & 76.51\%          & 14.29\%          & 6.03\%           \\
 \quad w/ views & \textbf{74.29\%} & \textbf{17.46\%} & \textbf{8.25\%}  & \textbf{23.49\%} \\ 
\noalign{\hrule height 1.2pt}
\end{tabular}
\end{center}
\vspace{-2em}
\end{table}

\section{Hierarchical Schema Linking}

\begin{figure}[ht]
  \includegraphics[width=0.485\textwidth]{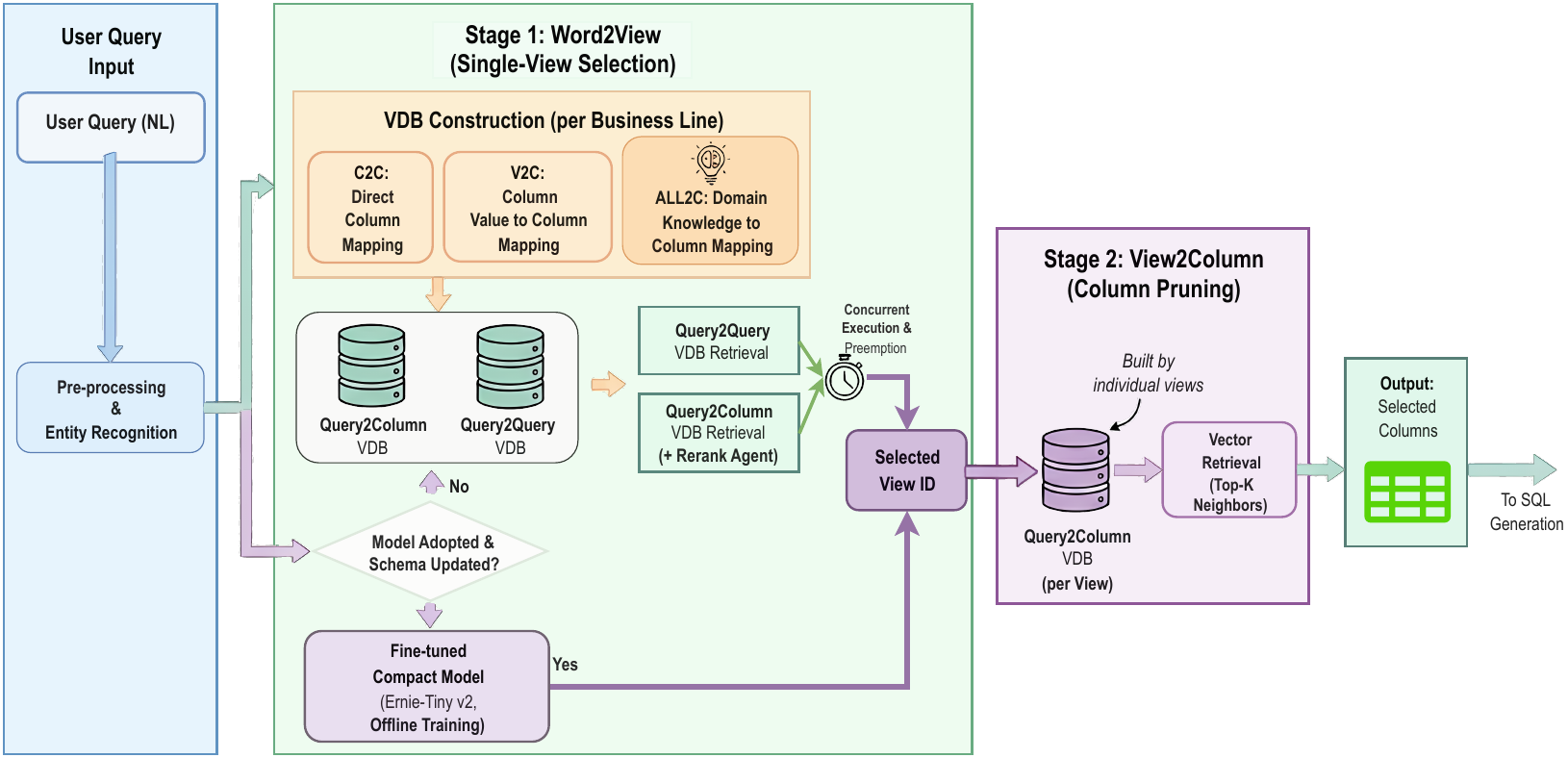}
  \caption{The Hierarchical Schema Linking module. }
  \label{figure:2stageMethod}
  \vspace{-1em}
\end{figure}

Figure~\ref{figure:2stageMethod} presents a detailed architecture of the Hierarchical Schema Linking module, which processes the natural language query of the user (following pre-processing and entity recognition) through two distinct stages: \textit{Word2View} and \textit{View2Column}.

In \textbf{Stage 1: Word2View (Single-View Selection)}, the system reformulates schema linking as a view routing problem. It relies on Vector Databases (VDBs) constructed per business line, which integrate three mapping strategies: Direct Column Mapping (C2C), Column Value to Column Mapping (V2C), and Domain Knowledge to Column Mapping (ALL2C). At runtime, the module dynamically determines the retrieval path. If a fine-tuned compact model (e.g., Ernie-Tiny v2) is adopted and the schema remains unchanged, it directly predicts the view. Otherwise, the system triggers concurrent execution between a Query2Column VDB retrieval (assisted by a rerank agent) and a Query2Query VDB retrieval, using preemption to minimize latency and output a \textbf{Selected View ID}.

In \textbf{Stage 2: View2Column (Column Pruning)}, the module refines the context by pruning irrelevant columns within the chosen view. Utilizing a dedicated Query2Column VDB built specifically for that individual view, it performs vector retrieval to extract the Top-K nearest neighbors. This stage outputs the final set of \textbf{Selected Columns}, which are subsequently forwarded to the SQL generation module.

\subsection{Word2View}

We propose a vector-retrieval approach to address the single-view selection problem. 
In the following, we first detail the three linking modes employed during the construction of the index of the vector database (VDB). 
We then introduce two retrievals: a Query2Query VDB retrieval tailored to report-oriented queries and a Query2Column VDB retrieval designed for ad-hoc queries. 
At runtime, both retrievals are executed concurrently, and the one that returns a result first preempts the other to meet targets for latency. 
Finally, we fine-tune a compact model offline on training data generated at this stage to further accelerate the selection of the single views. 

\subsubsection{Three Linking Modes}

During the construction of the index of the vector database, we introduce three linking modes—Direct Column Mapping (C2C), Column Value to Column Mapping (V2C), and Domain Knowledge to Column Mapping (ALL2C)—to address the core challenge of mapping natural language to schema elements. 
The first two mirror the modeling of SiriusBI~\cite{jiang2024siriusbi} concerning `Belongs-to' relationships at the value to column, column to table, and table to database levels. 
In contrast, ALL2C embeds domain knowledge directly into the indexing process to enhance the accuracy of schema linking.

The three linking modes are presented below. 
(a) C2C: the query entity corresponds to the column itself. 
We enumerate columns to generate seed queries, apply entity recognition to extract column mentions, and insert the embeddings of these mentions (from an embedding model, Embedding-V1~\cite{embedding-v1}) as keys into the Query2Column VDB, with the column identifier as the value. 
(b) V2C: the query entity is an enumerated value of a column. 
For example, a ``province'' column may have values such as `Sichuan', `Shanghai', or `Chongqing'. 
We enumerate such values, compute embeddings for them, and store them in the Query2Column VDB keyed by the value embedding with the associated column as the value. 
This enables queries like ``What is the DAU in Sichuan?'' to resolve to the ``province'' column via the enumerated value. 
(c) ALL2C: domain knowledge often captures common query paraphrases that implicitly refer to columns and values. 
For instance, ``the PV of Baidu'' in the advertising business line may be clarified as ``the page views for the advertiser Baidu'', where ``advertiser'' is a column, ``Baidu'' is one of the enumerated values of the column, and ``page views'' explains PV. 
Because the original phrasing may not directly reveal the target column, we store embeddings of such domain-derived paraphrases in the VDB and map them to the same corresponding column, thereby implementing synonym-based resolution. 

\subsubsection{Query2Column VDB Retrieval}

The \textit{Query2Column VDB} is built by business line and, within each business line, constructs a unified index over all views using three linking modes. 
Upon the arrival of a query, we first apply rule-based pre-processing to remove dimensions, such as dates, that can confound entity recognition. 
We then perform entity recognition on the sanitized query and feed the embeddings of the extracted entities, computed by an embedding model, as entry points for vector retrieval. 
Because identical columns may appear across multiple views, a rerank agent reorders the retrieved columns to identify a single view that best satisfies the query. 
This retrieval proves robust for ad-hoc queries with diverse phrasings, while the retrieval process continuously yields $<$query, ViewID$>$ pairs as training data.

\subsubsection{Query2Query VDB Retrieval}

To address the prevalence of recurring report queries in production, which—after rule-based normalization—exhibit highly similar forms and typically vary only along dimensions such as date, we propose the \textit{Query2Query VDB}. 
In this vector database, the key is the embedding of the normalized query after entity recognition, computed by an embedding model, and the value is the corresponding View ID. 
When the cosine similarity of the retrieved result exceeds $\alpha$, the query is deemed a match to a historical one.
Accordingly, the associated view ID is selected and returned as the Query2Query VDB retrieval result. 
This retrieval is well-suited to fixed-form recurring report queries and, relative to the Query2Column VDB retrieval, significantly reduces execution latency by minimizing reliance on the rerank agent. 
Moreover, the retrieval continuously generates training pairs of the form $<$query, ViewID$>$. 

\subsubsection{Fine-tuning a Compact Model}

We leverage the continuously accumulated $<$query, View ID$>$ pairs from retrieval to fine-tune a compact model. 
The model takes as input the embedding of each query after rule-based normalization and entity recognition, and outputs the corresponding View ID. 
We instantiate the backbone with Ernie-Tiny v2~\cite{sun2019ernie} (5.99M parameters). 
Training is conducted offline, and the newly fine-tuned model replaces the prior version upon completion. 
At inference, for each incoming query we first determine whether the relevant business line has adopted the fine-tuned model.
If not, we default to the \textit{Word2View} stage. 
If adopted, we compare the latest modification time of the schema metadata of the line—frequently updated in BI environments—with the training completion time of the compact model to decide whether to use the fine-tuned path. 
The selection of the schema linking strategy is transparent to users. 

\subsection{View2Column}

At this stage, we likewise use \textit{Query2Column VDB} for vector retrieval. 
Unlike the \textit{Word2View} stage, where the \textit{Query2Column VDB} is built by business lines, the \textit{View2Column} stage builds the \textit{Query2Column VDB} by individual views, constructing independent indices for each view using the three linking modes. 
This design effectively mitigates the challenge of excessive average column counts per view commonly encountered in real-world BI settings (as shown in Table~\ref{table1}). 
We return the Top-K nearest neighbors from the vector retrieval rather than only the highest-scoring neighbors. 
In views generated by the View Generation Agent, many columns exhibit similar query semantics (e.g., shared dimensions).
The incorporation of these additional columns as candidates enhances the accuracy of the SQL generation. 
We employ Mochow~\cite{mochow}, the in-house vector database of Baidu, as the underlying vector database. Preliminary evaluation demonstrates that the vector database seamlessly scales to thousands of views and tens of thousands of columns, maintaining a typical Recall@K metric of 99.1\% during the column retrieval phase. 

\subsection{Handling Schema Evolution}
To accommodate frequent schema changes, the system monitors metadata modifications and triggers asynchronous updates based on predefined thresholds. Rather than rebuilding the entire index, the system adopts an incremental update strategy to maintain efficiency. Specifically, the View Generation Agent updates relevant definitions, and the schema linking module inserts new embeddings without reconstructing the approximate nearest neighbor index. For deleted columns, the system masks the associated entities and applies tombstone flags within the vector database. Full rebuilds are reserved exclusively for the creation of new thematic views. This incremental approach ensures service continuity during updates. Moreover, the system prevents silent failures from stale views by returning null values for logically deleted columns and routing queries that target newly added but unindexed columns to the aforementioned expert interface.

\section{Routing Agent and the NL2IR2SQL Pipeline}\label{sec:routingAgent}

\begin{figure}[ht]
  \includegraphics[width=0.485\textwidth]{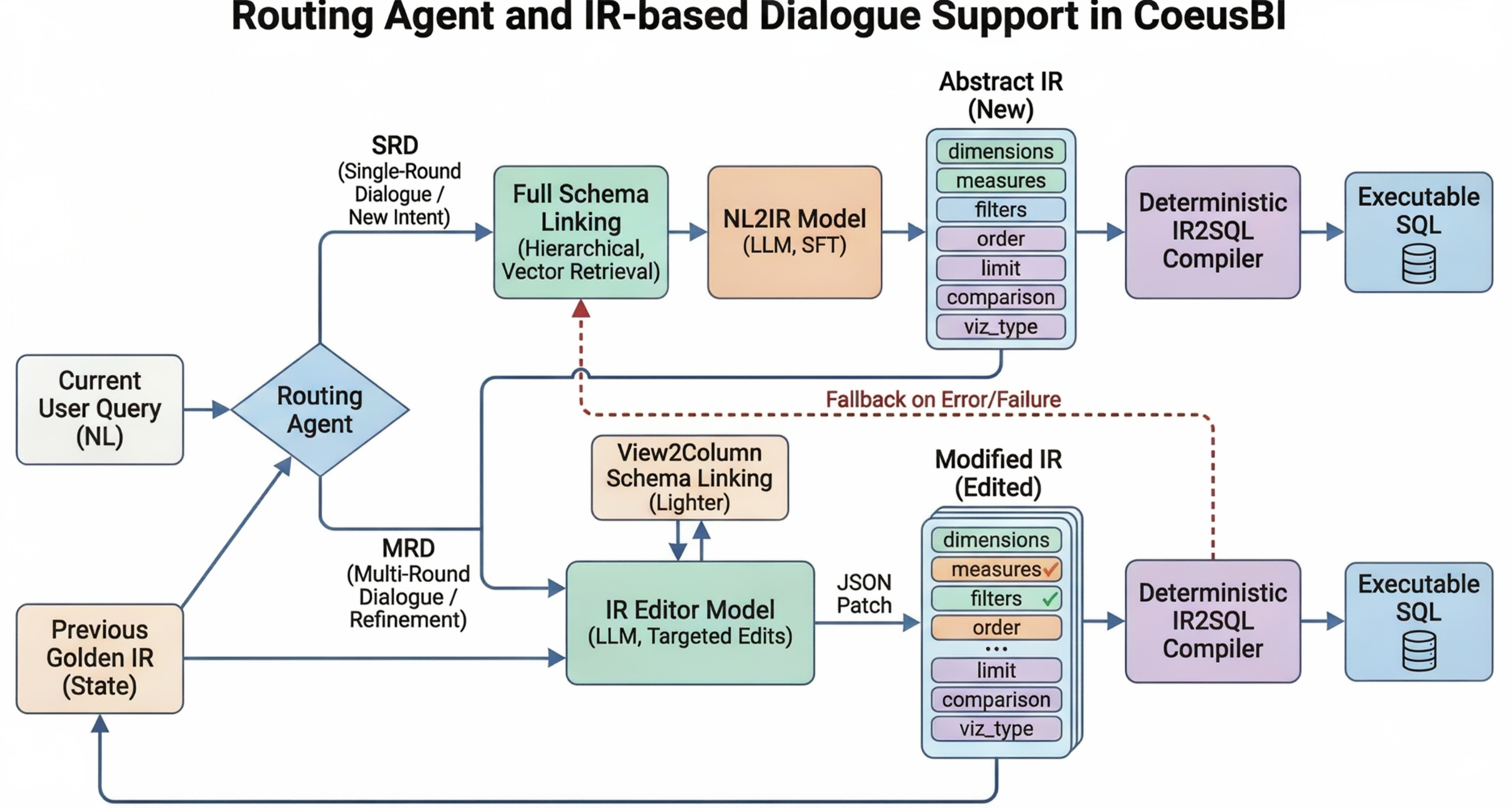}
  \caption{The Architecture and Pipeline of Routing Agent and IR-based Dialogue Support.}
  \label{figure:overviewOfRouting}
  % \vspace{-1em}
\end{figure}

To address the complexities of dialect support and conversational continuity, CoeusBI integrates a \textbf{Routing Agent} (as shown in Figure~\ref{figure:overviewOfRouting}) that acts as an intelligent controller for all incoming queries. 
Upon the arrival of a query, the Routing Agent evaluates the current natural language request alongside the previous golden Intermediate Representation (IR) state to determine whether the request constitutes an ongoing MRD refinement or a Single-Round Dialogue (SRD) requiring a novel intent. 

Based on this contextual judgment, the Routing Agent bifurcates the execution workflow:
\begin{itemize}
    \item \textbf{SRD Path (New Intent):} The query is routed through a full Hierarchical Schema Linking module (utilizing vector retrieval) and is processed by a fine-tuned NL2IR model based on LLMs to synthesize a new abstract IR.
    \item \textbf{MRD Path (Refinement):} The query is directed to a lighter \textit{View2Column} schema linking stage. An IR editor based on LLMs then evaluates the new constraints and applies targeted modifications via the JSON Patch protocol directly to the previous golden IR. 
\end{itemize}

Both paths ultimately converge on a deterministic compiler that translates the final dialect-agnostic IR into executable SQL. 
Furthermore, the IR is strictly separated from the view ID by design. 
The IR serves exclusively as the semantic generation target for the NL2IR model. 
The view ID, conversely, is an internal system identifier that the Word2View stage produces dynamically. 
To prevent the mixing of semantic comprehension with system-level metadata, the IR generated by the model deliberately excludes the view ID. 
Instead, the service layer maintains the selected view ID. 
To ensure robustness, a fallback safeguard is embedded within the pipeline: if an edited MRD representation fails to compile or execute, the system automatically redirects the query back to the SRD path (full schema linking and NL2IR) to prevent cascading conversational errors. 

Drawing on BI best practices, we adopt three core design principles for this overall pipeline: 1) use an IR that is agnostic to dialects; 2) restrict the IR to a well-defined subset of BI queries; and 3) provide deterministic compilation. 

\subsection{NL2IR: Structured, Dialect-Agnostic Mapping}

\begin{figure}[ht]
  \includegraphics[width=0.485\textwidth]{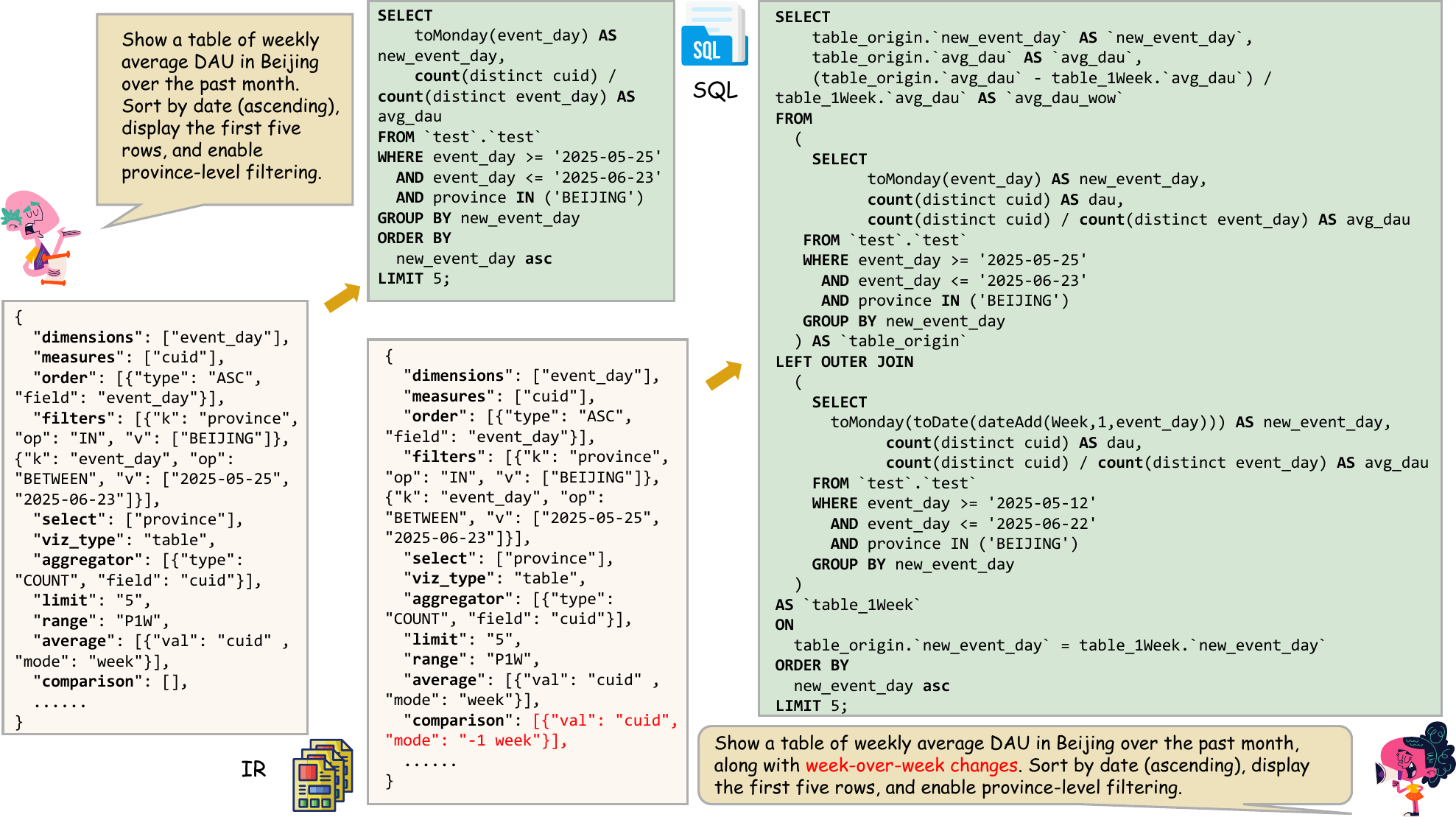}
  \caption{The Example of IR2SQL.}
  \label{figure:dsl2sql}
  \vspace{-1.5em}
\end{figure}

\subsubsection{Abstract Intermediate Representation}

As shown in Figure~\ref{figure:dsl2sql}, we design an abstract IR distilled from common SQL patterns in BI practice, covering the capabilities needed for typical BI queries.  
Moreover, this general IR is decoupled from database dialects, enabling unified support for multi-dialects via a single representation.
These fields in the IR are presented below: 

\begin{itemize}
\item {} \textit{dimensions.} Columns used for grouping in the result set, typically corresponding to GROUP BY in SQL. 
\item {} \textit{measures.} Core metrics such as UV, PV, DAU or uid, which represent the values to be aggregated or analyzed. 
\item {} \textit{order.} Sorting instructions, where field specifies the column to sort by and type indicates the direction. 
\item {} \textit{filters.} Conditions applied to filter the dataset, aligned with the WHERE clause in SQL (e.g., filtering by city = ``Beijing'').  
\item {} \textit{aggregator.} The aggregation function used on a given measure (e.g., SUM, AVG, COUNT). 
\item {} \textit{range.} A time-related constraint, such as P1M (past one month) or P1W (past one week), representing the temporal scope. 
\item {} \textit{average.} Indicates the need to compute an average on a specific measure over a defined period or dimension (e.g., weekly average). 
\item {} \textit{comparison.} Represents comparative logic such as week-over-week or year-over-year analysis, commonly used in BI reporting.  
\item {} \textit{limit.} A hard limit on the number of returned records. 
\item {} \textit{viz\_type.} Specifies the type of visualization (e.g., table, line chart, pie chart). 
\item {} \textit{select.} Indicates whether a global filter component should be enabled for a given column, supporting interactive visual analysis.
\end{itemize}
Owing to space constraints, features such as time-over-time comparisons, proportional analyses, and aggregations across different dimensions are not detailed here. 
Because the IR constrains structure and semantics to common BI patterns, NL2IR reduces ambiguity and the effective search space relative to direct NL2SQL, which empirically yields higher generation accuracy (more details are discussed in Section~\ref{evaluation:ab1}). 

Beyond support for multiple dialects, we introduce a viz\_type field in the IR to satisfy visualization requirements in BI workloads, enabling line, bar, and pie chart renderings. 
As shown in Figure~\ref{figure:dsl2sql}, common period-over-period analyses (e.g., year-over-year or month-over-month), which correspond to nested queries, can be readily realized by specifying the relevant dimension within the \textit{comparison} capability. 

\subsubsection{Training Methodology}

Constrained by data privacy considerations, we employ the \textit{DeepSeek-R1-Distill-Qwen-32B} model and conduct SFT to serve as the NL2IR converter in the production environment.
This is accomplished using two nodes equipped with a total of 16 NVIDIA A800-80GB GPUs.
The SFT process is configured for 3 epochs, with a learning rate of 3e-5, a global batch size of 16, and a sequence length of 32768 tokens.
To ensure fair comparisons during the evaluations, the experiments omit this production model. Instead, the evaluations employ the identical backbone models (such as DeepSeek-V3 or XiYanSQL-QwenCoder-32B-2504 (named as ``QwenCoder-2504'' in our paper)) utilized by the baseline methods across all components requiring LLMs, including the NL2IR converter and the IR editor.
We make the instruction prompts for SFT publicly available at \href{https://github.com/rebornDanny/CoeusBI}{githubRepo}.

\subsection{Deterministic IR2SQL Compilation and Coverage Guarantees} 

We implement a deterministic IR-to-SQL compiler: for the covered BI subset, compilation is predictable and preserves the intended semantics. 
Figure~\ref{figure:dsl2sql} illustrates a concrete example. 
In the first query, the user requests the weekly average ``DAU'' for Beijing over the most recent month, sorted by date in ascending order, displayed as the first five rows in a tabular view, with a province-level filter enabled. 
In the second query, the user additionally requests the week-over-week (WoW) comparison. 
While the corresponding SQL must employ a nested subquery to encode this comparative semantics, in the IR it suffices to augment the \textit{comparison} field with a ``-1 Week'' offset. 

\subsubsection{Virtual Columns and Error Handling}

To expand system applicability and address scenarios where the IR cannot natively express advanced SQL features, we introduce \textit{Virtual Columns}. By pushing computational complexities like nested subqueries, advanced window functions, and specialized deduplication logic down into the semantic view layer, this mechanism decouples complex business logic from IR syntax. The IR simply treats these \textit{virtual columns} as standard measures or dimensions. This architectural separation ensures the front-end IR remains lightweight, deterministic, and easily generated by the LLM, while the underlying view absorbs domain-specific SQL complexity. Operationally, data experts periodically review failed queries to define \textit{Virtual Columns} for patterns the IR cannot capture or routinely misrepresents. For example, in video analytics, uid denotes the logged-in user identifier, whereas the daily active users (DAU) metric is not stored directly and must be computed. Because LLMs might omit the DISTINCT keyword during generation, yielding erroneous results from duplicate records, we create a DAU Virtual Column within the view with a computation rule like ``count(distinct if(uid is not null, uid, null))'', adapted for the target database dialect. This enforces the use of distinct in all subsequent DAU related queries, thereby improving generation accuracy.

\subsection{Expressiveness and Limitations of the Representation}

While the intermediate representation is highly effective for typical analytical workloads, the bounded expressiveness of the representation is a deliberate architectural design choice rather than an unintentional limitation. In enterprise business intelligence environments, guaranteeing deterministic compilation and eliminating hallucination risks associated with LLMs for the vast majority of production queries strictly take priority over unconstrained SQL expressiveness. By design, the IR restricts its scope to a well-defined subset of business queries to safely remove non-deterministic generation from the final SQL synthesis step.

Currently, the representation does not fully support complex nested subqueries (beyond predefined comparative logic), advanced window functions (e.g., recursive ranking or moving averages), or non-standard groupings. However, the system mitigates this limitation through the combination of a strictly bounded IR and expressive virtual columns. Consequently, highly specialized queries outside the conventional business reporting scope may fail to map accurately. During the production deployment documented in this study, the single-view IR pipeline successfully expresses and executes 95.3\% of user queries end-to-end. The remaining 4.7\% of queries fail to process autonomously. These failures originate from out-of-coverage intents, such as queries requiring complex cross-view logic, and pipeline errors, including schema-linking misclassifications and infrastructure faults. To address these cases, the system routes the failed queries to the expert fallback mechanism described in Section~\ref{sec:overview}. The View Generation Agent subsequently utilizes the accumulated feedback to automatically construct new single views and virtual columns, continuously expanding the functional coverage of the single-view representation without introducing the hallucination vulnerabilities associated with free-form SQL generation. Future work will focus on expanding the semantics of the representation to encompass these advanced SQL features without compromising deterministic compilation. 

\section{IR-based Dialogue Support}

Building upon the routing logic and fallback mechanisms detailed in Section~\ref{sec:routingAgent}, we directly present a concrete conversational sequence from the production environment of Baidu to illustrate the practical application of the SRD and MRD paths.

\begin{figure}[ht]
  \includegraphics[width=0.485\textwidth]{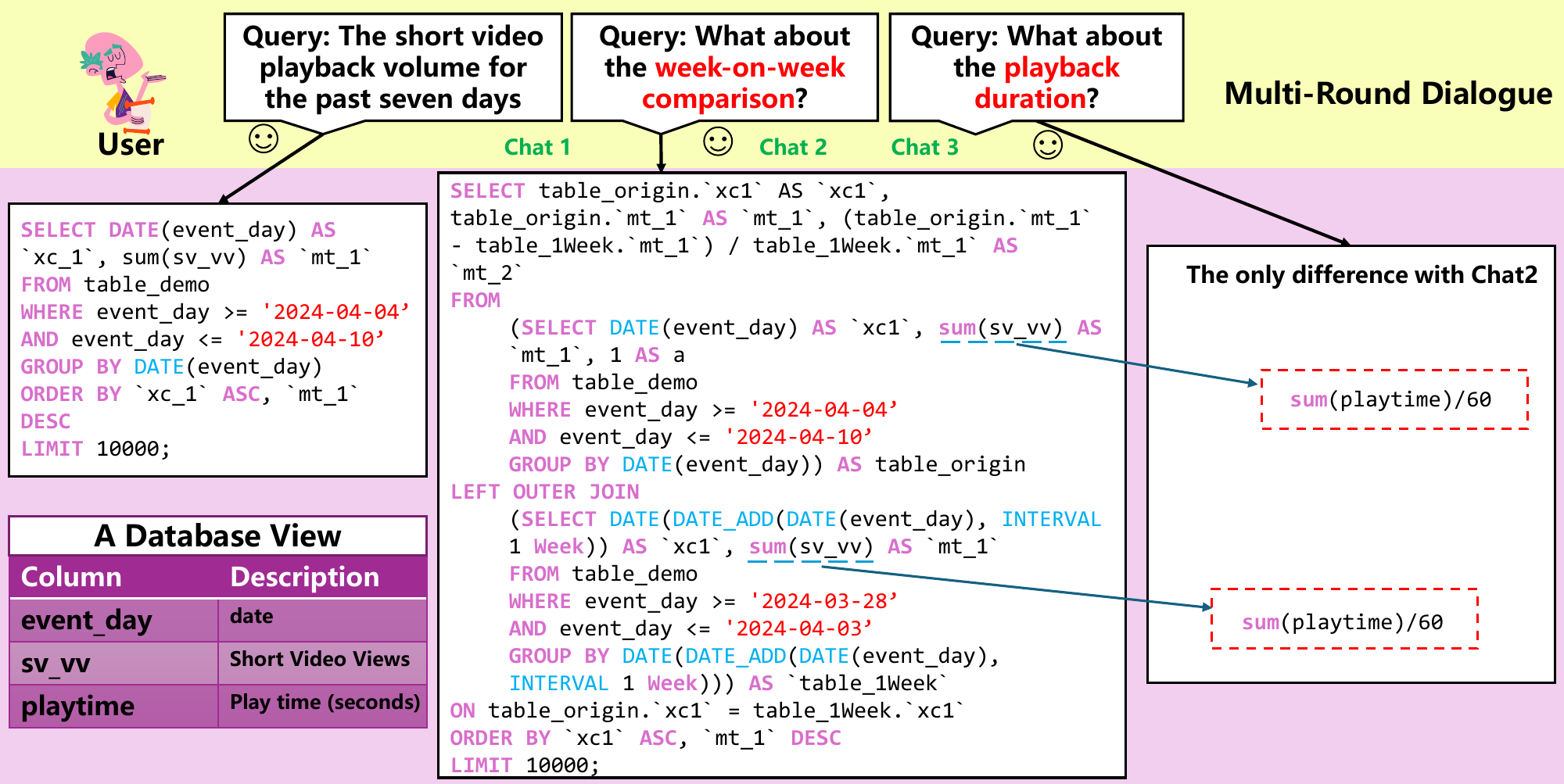}
  \caption{The Example of an MRD in Baidu. }
  \label{figure:exampleOfMRD}
  \vspace{-1em}
\end{figure}

As shown in Figure~\ref{figure:exampleOfMRD}, a user issues three queries in Baidu (`Chat1', `Chat2', and `Chat3'), with the latter two forming an MRD. 
For `Chat1', there is no previous IR, so the Routing Agent classifies the query as SRD and invokes the Hierarchical Schema Linking module to obtain the linking result. 
The query is then processed by the NL2IR2SQL pipeline: the natural language query is converted into the corresponding IR, which is compiled into executable SQL and the final analytical output. 

For `Chat2', whose intent is ``What is the short video playback volume for the past seven days and its week-on-week comparison?'', the agent uses the current query together with the golden IR of `Chat1' and classifies the query as MRD. 
Unlike `Chat1', which undergoes the full schema linking and NL2IR processes, `Chat2' only requires editing the golden IR derived from `Chat1'. 
As shown in Figure~\ref{figure:dsl2sql}, the intent is satisfied by adding ``-1 week'' to the comparison field within the IR.
The modified IR then serves as the golden IR of Chat2 to support subsequent MRD. 

For `Chat3', whose intent is ``What is the short video playback duration for the past seven days and its week-on-week comparison?'', the query is likewise classified as an MRD. 
Unlike `Chat2', this query requires schema linking for the changed column. 
Rather than executing the full schema linking process as in `Chat1', the Routing Agent leverages the view ID associated with the golden IR at the service layer to undergo the lighter \textit{View2Column} stage for schema linking. 
Consequently, the full NL2IR process is not invoked.
The outcome of schema linking is applied to modify the golden IR via the JSON Patch protocol~\cite{jsonPatch}. 
In this example, the IR for `Chat3' differs from the golden IR only by replacing the sv\_vv column with playtime in the \textit{measures} field.

\textbf{Error Handling.}
Error handling in the MRD support module requires high view generation quality, particularly when schema linking invokes only the View2Column stage. Production dataset evaluations show the Routing Agent achieves 98.2\% classification accuracy for SRD queries and 98.6\% for MRD queries. Misclassifications create operational challenges. Classifying an MRD query as an SRD query discards historical context, generating a malformed intermediate representation that triggers a hard compilation failure. Conversely, treating an SRD query as an MRD query reuses the previous view, causing a hard failure if required metrics are absent, or a silent semantic failure (an executable but incorrect result) if identically named columns possess different meanings. To mitigate hard failures, as outlined in Section~\ref{sec:routingAgent}, the system automatically redirects failed queries through the full SRD pipeline. If this fallback fails, queries are logged and processed through a centralized expert feedback cycle to refine views based on actual conversational patterns. To address silent semantic failures without human validation latency, a transparent user interface explicitly displays resolved dimensions and measures for rapid semantic verification. Upon observing discrepancies, users can trigger a retry prompting a self-contained query, thereby breaking the misclassification chain and ensuring high throughput.

As demonstrated by the preceding interaction sequence, this architecture intrinsically facilitates efficient session management. By maintaining only the selected view identifier and the immediate predecessor's intermediate representation, the system avoids the storage and processing overhead of appending complete dialogue histories. Operating the multi-round editor on a fixed-size context ensures constant token consumption and inference latency during extended exploratory dialogues, preventing performance degradation as session length increases.

\section{Experimental Evaluation}\label{section:experiment}

\begin{figure}[ht]
  \includegraphics[width=0.485\textwidth]{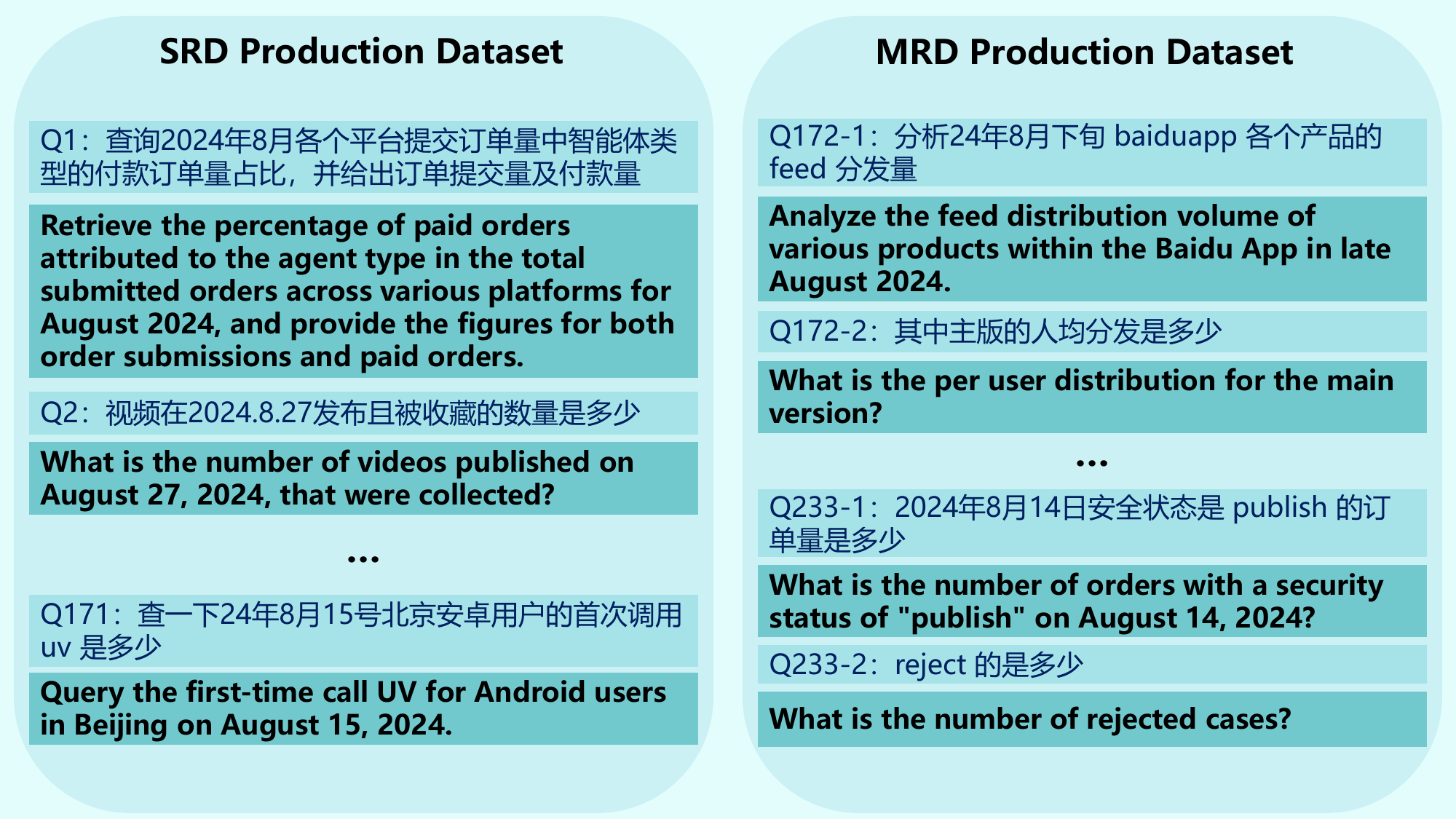}
  \caption{SRD production dataset and MRD production dataset.}
  \label{figure:Dataset}
  \vspace{-1.5em}
\end{figure}

To evaluate our framework, in this section, we provide a detailed analysis from three perspectives: 
1) we conduct End-to-End (E2E) evaluations of CoeusBI on both public datasets and production datasets to demonstrate overall system performance; 
2) we validate the effectiveness of each module with experiments targeted to the specific subproblems the module is designed to address; 
3) we perform ablation studies to isolate and quantify the contribution of each module to the overall performance. 

\subsection{Experiment Setups}

\subsubsection{Datasets} 
To comprehensively evaluate the overall performance of CoeusBI and the effectiveness of individual modules, we conduct extensive experiments across multiple tasks using both public datasets and production datasets. 
Due to data privacy concerns, we release partially anonymized versions of two production datasets, as well as the generated views for the public datasets, which are available at \href{https://github.com/rebornDanny/CoeusBI}{githubRepo}. 

\textbf{SRD Datasets}
To evaluate the effectiveness of the SQL generation module, we adopt BIRD~\cite{li2024can}, a challenging large-scale NL2SQL dataset designed to bridge the gap between academic research and applications in the real world. 
BIRD includes 95 large databases and features high-quality pairs of user questions and SQL queries across 37 domains. 
In addition to BIRD, we collect 171 SRD queries from four different business lines at Baidu that use ClickHouse~\cite{ClickHouse}, encompassing 148 tables and 25,258 columns, to construct a dataset referred to as the SRD production dataset. 

\textbf{MRD Dataset}
Similarly, following the same logic as the SRD production dataset, we construct the MRD production dataset at Baidu, which runs on ClickHouse and comprises 62 query sets, each spanning two to three dialogue rounds, for a total of 144 queries. 
Figure~\ref{figure:Dataset} shows examples of both production datasets. 

\subsubsection{Evaluation Metrics}

We evaluate using five metrics overall: three effectiveness metrics—\textit{Execution Accuracy} (EX), \textit{Valid Efficiency Score} (VES), and \textit{Useful Execution Accuracy} (UEX)—and two cost metrics—\textit{prompt tokens} (PT) and \textit{response tokens} (RT). 
EX relies solely on execution outcomes relative to the golden SQL queries of the dataset, which can lead to misjudgments. 
For instance, for the BI query ``\textit{Yesterday's interaction metrics for users of app1}'', the golden SQL is:
\begin{lstlisting}[ language=SQL,
                    deletekeywords={IDENTITY},
                    deletekeywords={[2]INT},
                    morekeywords={clustered},
                    framesep=8pt,
                    xleftmargin=40pt,
                    framexleftmargin=40pt,
                    frame=tb,
                    framerule=0pt ]
SELECT SUM(click_pv) AS total_click_pv, 
     SUM(disp_pv) AS total_disp_pv, 
     SUM(play_pv) AS total_play_pv, 
     SUM(like_pv) AS total_like_pv, 
     SUM(comment_pv) AS total_comment_pv, 
     SUM(collection_pv) AS total_collection_pv, 
     SUM(shareto_pv) AS total_shareto_pv, 
     SUM(follow_pv) AS total_follow_pv
FROM table1
WHERE event_day = DATE_SUB(CURRENT_DATE, INTERVAL 1 DAY) 
AND appid = 'app1';
\end{lstlisting}
These columns (click\_pv, disp\_pv, play\_pv, like\_pv, comment\_pv, collection\_pv, shareto\_pv, follow\_pv) are interaction metrics.
Yet any non-empty subset of these columns can be considered correct for the intent. 
The following SQL is therefore also deemed correct: 
\begin{lstlisting}[ language=SQL,
                    deletekeywords={IDENTITY},
                    deletekeywords={[2]INT},
                    morekeywords={clustered},
                    framesep=8pt,
                    xleftmargin=40pt,
                    framexleftmargin=40pt,
                    frame=tb,
                    framerule=0pt ]
SELECT SUM(click_pv) AS total_click_pv, 
     SUM(disp_pv) AS total_disp_pv
FROM table1
WHERE event_day = DATE_SUB(CURRENT_DATE, INTERVAL 1 DAY) 
AND appid = 'app1';
\end{lstlisting}
Following prior work~\cite{floratou2024nl2sql, lei2024spider}, we adopt \textit{usefulness} rather than strict \textit{correctness} as the criterion. 
Thus, we use the UEX metric, which assesses whether the execution output of the generated SQL aligns with the query intent. 
Specifically, since both datasets include golden SQL queries, we first perform an EX match to evaluate all results. 
Any result initially marked incorrect is then reviewed by a panel of three database administrators (DBAs). 
If at least two out of three DBAs determine that the query result aligns with the intent and should be considered correct, the label is updated accordingly. 
Corrections are applied only in the following three cases: 1) when the golden SQL omits an aggregation specification and the difference in outcomes between the golden SQL and the checked SQL arises solely from the choice of aggregation; and 2) when the columns selected by the golden SQL constitute a subset of those selected by the checked SQL (i.e., the latter includes additional columns), and the extra columns do not affect the analytical result; 
and 3) when the actual queried results fundamentally satisfy the underlying user intent despite structural deviations from the golden SQL. 
For NL2SQL tasks on the BIRD dataset, we adhere to the standard EX and VES as outlined in~\cite{li2024can}. 
For NL2BI tasks, we employ UEX as the effectiveness metric and report PT and RT to assess the economic cost.

\subsubsection{Baselines and Implementation Details}
We evaluate the NL2BI capability of CoeusBI by comparing it against four baseline systems: 1) DIN-SQL~\cite{pourreza2024din}, 2) MAC-SQL~\cite{wang2025mac}, 3) Xiyan-SQL~\cite{gao2024xiyan} (all oriented toward NL2SQL) and 4) SiriusBI~\cite{jiang2024siriusbi} (oriented toward NL2BI). 
We omit other commercial or manual-heavy systems discussed in the related work (as same as SiriusBI~\cite{jiang2024siriusbi}) because: 1) non-interactive tools structurally misalign with our conversational BI focus; 2) manual semantic modeling (e.g., via LookML) introduces human bias, rendering comparisons with our automated system unfair; and 3) strict data privacy regulations prohibit uploading production datasets to external platforms.
For all baseline methods, we adhere closely to published configurations, including model architectures and prompt designs. 
The original implementations of DIN-SQL, MAC-SQL, and Xiyan-SQL are designed for SRD queries and lack native support for MRDs. 
Therefore, we extend the prompts of these models with minimal modifications to ensure compatibility with the MRD production dataset while preserving core functionality.
The modified prompts are available at \href{https://github.com/rebornDanny/CoeusBI}{githubRepo}. 
For evaluations on the BIRD-dev dataset, the baseline executions of DIN-SQL and MAC-SQL utilize GPT-4o as the underlying model~\cite{jiang2024siriusbi}. 
For fair comparison, we evaluate Xiyan-SQL and SiriusBI using XiYanSQL-QwenCoder-2504~\cite{gao2024xiyan}, following the recommended configurations of the authors. 
In CoeusBI, the parameter K in the View2Column stage is 50 and $\alpha$ is 0.95.

\subsection{End-to-End Performance}

In this experiment, we evaluate the E2E performance of various baselines on the SRD and MRD datasets.

\subsubsection{Results on BIRD-dev Dataset} 

\begin{table}[htbp]
\begin{center}
\caption{Evaluations on the BIRD-dev dataset. We chose GPT-4o for the same reason as in the SiriusBI experiment - we select the same backbone.}
\label{exp:bird-dev}
\begin{tabular}{lcc}
\noalign{\hrule height 1.2pt}
Method         & EX (\%)        & VES (\%) \\ \hline
DIN-SQL + GPT-4o          & 50.72          & 58.79    \\
DIN-SQL + DeepSeek-V3     & 59.71           & 65.85     \\
MAC-SQL + GPT-4o          & 57.56          & 58.04    \\
MAC-SQL + DeepSeek-V3          & 60.10          & 65.33    \\
SiriusBI-QwenCoder-2504         & 68.97          & 70.89    \\
Xiyan-SQL-QwenCoder-2504        & \textbf{73.34} & --       \\
CoeusBI + DeepSeek-V3 & 71.4          & 72.19    \\
CoeusBI-QwenCoder-2504 & 70.6          & 71.3     \\
\noalign{\hrule height 1.2pt}
\end{tabular}
\end{center}
% \vspace{-1.5em}
\end{table}

\paragraph{Superiority Through Search Space Reduction} Table~\ref{exp:bird-dev} demonstrates that when employing the DeepSeek-V3 backbone, CoeusBI achieves an EX of 71.4\% and a VES of 72.19\%, significantly outperforming the general DIN-SQL framework which yields an EX of 59.7\%. This substantial improvement indicates that the specialized intermediate representation and view generation mechanisms of CoeusBI effectively restrict the search space compared to free-form generation pipelines. Furthermore, this architectural advantage persists against industrial baselines, as CoeusBI utilizing QwenCoder-2504 attains an EX of 70.6\%, maintaining a clear edge over the 68.97\% achieved by the state-of-the-art method SiriusBI.

\subsubsection{Results on both Production Datasets} 

\begin{table}[htbp]
\begin{center}
\caption{Evaluations on both production datasets.}
\label{exp:SRDDataset}
\begin{tabular}{lccc}
\noalign{\hrule height 1.2pt}
Method                                        & UEX (\%)      & PT              & RT              \\ \hline
\multicolumn{4}{c}{SRD production dataset (ClickHouse)}                                           \\ \hline
DIN-SQL + DeepSeek-V3                           & 28.7          & 2948803         & 109467          \\
MAC-SQL + DeepSeek-V3                           & 59.1          & 3645471         & 87594           \\
SiriusBI-QwenCoder-2504                         & 57.8          & 2943395         & 19446           \\
XiYanSQL-QwenCoder-2504                         & 42.7          & 1557352         & \textbf{10987}  \\
CoeusBI + DeepSeek-V3          & \textbf{86.5}          & 806769          & 31446           \\ 
CoeusBI-QwenCoder-2504       & 84.2  & \textbf{519623} & 31402 \\ \hline
\multicolumn{4}{c}{MRD production dataset (ClickHouse)}                                           \\ \hline
DIN-SQL + DeepSeek-V3                           & 16.0          & 2521508         & 147633          \\
MAC-SQL + DeepSeek-V3                           & 43.8          & 2540791         & 73500           \\
SiriusBI-QwenCoder-2504                          & 56.9          & 3478433         & 47258           \\
XiYanSQL-QwenCoder-2504                         & 28.5          & 1656397         & \textbf{25966}  \\
CoeusBI + DeepSeek-V3          & \textbf{83.3}          & 683025          & 31363           \\ 
CoeusBI-QwenCoder-2504       & 82.6  & \textbf{461901} & 38492 \\ 
\noalign{\hrule height 1.2pt}
\end{tabular}
\end{center}
\vspace{-1.5em}
\end{table}

\paragraph{Dominant Execution Accuracy in Real-World Scenarios} The evaluations on the SRD and MRD production datasets reveal that CoeusBI achieves unprecedented execution accuracy compared to existing frameworks. While industrial baselines like SiriusBI max out at 57.8\% and 56.9\% UEX on the SRD and MRD datasets respectively, CoeusBI equipped with DeepSeek-V3 establishes a new state-of-the-art performance of 86.5\% and 83.3\%. Notably, XiYanSQL, which previously demonstrated strong academic performance, experiences severe performance degradation on these ClickHouse-based production datasets, dropping to 42.7\% and 28.5\% UEX. This stark contrast highlights the structural robustness of CoeusBI and its superior capability to handle the domain-specific complexities inherent in real-world industrial databases.

\paragraph{Exceptional Prompt Token Efficiency} Beyond execution accuracy, CoeusBI exhibits a massive reduction in computational overhead as measured by prompt tokens. Traditional frameworks such as MAC-SQL and SiriusBI consume excessive prompt tokens, typically ranging from 2.5 million to over 3.6 million across the two datasets. In contrast, the CoeusBI framework operates with remarkable conciseness. The QwenCoder-2504 variant of CoeusBI achieves the lowest prompt token counts of 519623 on SRD and 461901 on MRD, representing an approximately 80\% to 85\% reduction compared to MAC-SQL. This structural efficiency proves that CoeusBI can synthesize highly effective context representations without relying on bloated prompts, drastically optimizing inference costs.

\paragraph{Optimal Balance of Generation Efficiency and Accuracy} An analysis of response tokens demonstrates that CoeusBI achieves an ideal trade-off between generation length and execution correctness. Although XiYanSQL yields the lowest response tokens across both datasets, this conciseness comes at the severe expense of UEX, rendering it impractical for robust production. CoeusBI, however, maintains highly competitive response token counts around 31000 while delivering peak accuracy. Furthermore, comparing the underlying models within the CoeusBI architecture reveals that transitioning from DeepSeek-V3 to QwenCoder-2504 results in a negligible UEX penalty of 2.3\% on SRD and 0.7\% on MRD, while further minimizing prompt tokens. This validates the QwenCoder-2504 configuration as a highly cost-effective and dependable alternative for resource-constrained deployments.

\subsection{Detailed Analysis}

\subsubsection{Evaluation of View Generation Agent}

To ensure reproducibility, we evaluate the View Generation Agent using the BIRD-dev dataset and the open-source DIN-SQL method.
Table~\ref{tableview} shows that the View Generation Agent effectively reduces the proportion of join operations in BIRD-dev, thereby improving the EX metric. 

\begin{table}[ht]
\caption{Evaluation of View Generation Agent on the student\_club dataset. }
\begin{center}
\label{tableview2}
\resizebox{\columnwidth}{!}{
\begin{tabular}{lcccc}
\noalign{\hrule height 1.2pt}
Method & Easy & Non-nested & Nested & EX \\ \hline
DIN-SQL + DeepSeek-R1-0528 & 20.75\% & 71.70\% & 7.50\% & 75.32\% \\
\quad w/ views w/o descriptions & 89.17\% & 3.80\% & \textbf{7.01\%} & 77.85\% \\
\quad w/ views w/ descriptions  & \textbf{89.87\%} & \textbf{1.90\%} & 8.23\% & \textbf{79.75\%} \\ 
\noalign{\hrule height 1.2pt}
\end{tabular}
}
\end{center}
% \vspace{-1.5em}
\end{table}

We further evaluate the extent to which views simplify query complexity and the impact of the descriptions on the generation of the views on the BIRD-dev student\_club database. 
Table~\ref{tableview2} shows that, without views, the classifier of DIN-SQL classifies 20.75\% of queries as \textit{Easy}, 71.70\% as \textit{Non-nested}, and 7.50\% as \textit{Nested}. 
Across all view-based variants, the share of \textit{Easy} queries rises to at least 89.17\%, while \textit{Non-nested} queries fall to at most 3.8\%, indicating that views substantially reduce query complexity. 
Both DIN-SQL with the views of CoeusBI without descriptions and DIN-SQL with the views of CoeusBI with descriptions improve execution accuracy by converting a larger fraction of queries that include JOIN clauses into \textit{Easy} queries. 
Moreover, DIN-SQL with the views of CoeusBI with descriptions outperforms DIN-SQL with the views of CoeusBI without descriptions, implying that the incorporation of descriptions positively influences the quality of the view generation. 

\begin{table}[ht]
\caption{Effect of materializing the auto-generated views on the execution accuracy of DIN-SQL and query execution time on the dataset of BD-Business Line A.}
\begin{center}
\label{tableview3}
\resizebox{\columnwidth}{!}{
\begin{tabular}{lccc}
\noalign{\hrule height 1.2pt}
Method                                                       & EX               & Time           & Memory          \\ \hline
DIN-SQL + DeepSeek-R1-0528                         & 6.03\%           & 317 s          & \textbf{301 GB} \\
\quad w/ Top-3 most-used views materialized & \textbf{23.49\%} & 221 s          & 302 GB          \\
\quad w/ Top-5 most-used views materialized & \textbf{23.49\%} & \textbf{189 s} & 303 GB          \\ 
\noalign{\hrule height 1.2pt}
\end{tabular}
}
\end{center}
% \vspace{-1.5em}
\end{table}

\begin{table*}[htbp]
    \centering
    \caption{Comprehensive comparison of manual and automated view generation on the student\_club dataset. (Due to layout length constraints, ``DeepSeek'' here refers to the DeepSeek-v4-pro backbone.)}
    \label{exp:manual_vs_auto}
    \vspace{0.2cm}
    \begin{subtable}{0.48\textwidth}
        \centering
        \caption{Execution Accuracy (EX \%)}
        \resizebox{\textwidth}{!}{
        \begin{tabular}{lccc}
        \noalign{\hrule height 1.2pt}
        View Source & GPT-4o & GPT-5.5 & DeepSeek \\
        \hline
        DeepSeek-v4-pro & 77.85 & 81.65 & 81.75 \\
        GPT-5.5 & 75.32 & 78.48 & 82.91 \\
        GPT-4o & 75.95 & 77.22 & 79.75 \\
        Manual & 76.58 & 77.22 & 80.38 \\
        No Views & 74.68 & 75.95 & 76.58 \\
        \noalign{\hrule height 1.2pt}
        \end{tabular}
        }
    \end{subtable}
    \hfill
    \begin{subtable}{0.48\textwidth}
        \centering
        \caption{Prompt Tokens (PT)}
        \resizebox{\textwidth}{!}{
        \begin{tabular}{lccc}
        \noalign{\hrule height 1.2pt}
        View Source & GPT-4o & GPT-5.5 & DeepSeek \\
        \hline
        DeepSeek-v4-pro & 5060836 & 5110781 & 5371351 \\
        GPT-5.5 & 5008571 & 5102610 & 5383453 \\
        GPT-4o & 5852557 & 5907074 & 6213743 \\
        Manual & 9286833 & 9411855 & 9995219 \\
        No Views & 3977146 & 3648173 & 3647804 \\
        \noalign{\hrule height 1.2pt}
        \end{tabular}
        }
    \end{subtable}
    \vspace{0.3cm}
    \begin{subtable}{0.48\textwidth}
        \centering
        \caption{Response Tokens (RT)}
        \resizebox{\textwidth}{!}{
        \begin{tabular}{lccc}
        \noalign{\hrule height 1.2pt}
        View Source & GPT-4o & GPT-5.5 & DeepSeek \\
        \hline
        DeepSeek-v4-pro & 125211 & 179413 & 375410 \\
        GPT-5.5 & 118174 & 180532 & 373689 \\
        GPT-4o & 128839 & 179798 & 414918 \\
        Manual & 123906 & 153741 & 367080 \\
        No Views & 126617 & 149075 & 323362 \\
        \noalign{\hrule height 1.2pt}
        \end{tabular}
        }
    \end{subtable}
    \hfill
    \begin{subtable}{0.48\textwidth}
        \centering
        \caption{Cost (USD \$)}
        \resizebox{\textwidth}{!}{
        \begin{tabular}{lccc}
        \noalign{\hrule height 1.2pt}
        View Source & GPT-4o & GPT-5.5 & DeepSeek \\
        \hline
        DeepSeek-v4-pro & 13.90 & 30.94 & 10.65 \\
        GPT-5.5 & 13.70 & 30.93 & 10.67 \\
        GPT-4o & 15.92 & 34.93 & 12.26 \\
        Manual & 24.46 & 51.67 & 18.67 \\
        No Views & 11.21 & 22.71 & 7.47 \\
        \noalign{\hrule height 1.2pt}
        \end{tabular}
        }
    \end{subtable}
    \vspace{0.3cm}
    \begin{subtable}{0.48\textwidth}
        \centering
        \caption{Ratio of EASY Queries (\%)}
        \resizebox{\textwidth}{!}{
        \begin{tabular}{lccc}
        \noalign{\hrule height 1.2pt}
        View Source & GPT-4o & GPT-5.5 & DeepSeek \\
        \hline
        DeepSeek-v4-pro & 34.81 & 84.18 & 77.21 \\
        GPT-5.5 & 47.47 & 96.84 & 90.51 \\
        GPT-4o & 43.67 & 73.42 & 65.82 \\
        Manual & 41.77 & 69.62 & 67.72 \\
        No Views & 22.78 & 22.15 & 22.78 \\
        \noalign{\hrule height 1.2pt}
        \end{tabular}
        }
    \end{subtable}
    \hfill
    \begin{subtable}{0.48\textwidth}
        \centering
        \caption{View Generation Time (Seconds)}
        \resizebox{0.55\textwidth}{!}{
        \begin{tabular}{lc}
        \noalign{\hrule height 1.2pt}
        View Source & Time (s) \\
        \hline
        DeepSeek-v4-pro & 116.4 \\
        GPT-5.5 & 47.62 \\
        GPT-4o & 29.2 \\
        Manual & 28689 \\
        No Views & 0 \\
        \noalign{\hrule height 1.2pt}
        \end{tabular}
        }
    \end{subtable}
\end{table*}

To evaluate the impact of this agent on query execution time, we employ materialized views on the BD-Business Line A dataset, which features large base tables. 
Table~\ref{tableview3} shows that materializing the top three and five most frequently used views reduces the average execution time of queries that include JOIN clauses from 317 seconds to 221 seconds and 189 seconds, respectively. 
The storage overhead incurred by materialized views is modest. 
Thus, this module enhances both the accuracy of SQL generation and the efficiency of query execution.

\subsubsection{Comparison with Manual Semantic Modeling}
To directly evaluate the efficacy of the automated view generation against manual semantic modeling, we conduct a controlled experiment on the student\_club database from the BIRD-dev dataset. The experiment isolates the view-authoring step and utilizes the external \textsc{DIN-SQL} framework as the common consumer of the generated views. Three data analysts manually construct semantic views under a blind setting. The final manual baseline constitutes a mixture of the views authored by these experts. Following the identical protocol, we evaluate the \textsc{DIN-SQL} pipeline across different view sources: automated generation by various language models, manual authoring, and a baseline without views. The evaluations employ GPT-4o, GPT-5.5, and DeepSeek-v4-pro as the backbone evaluators of the \textsc{DIN-SQL} pipeline to verify robustness.

Table~\ref{exp:manual_vs_auto} presents the comprehensive comparison. The automated view generation achieves execution accuracy that is comparable to or slightly higher than the accuracy of manual semantic modeling. Specifically, the strongest automated agent, DeepSeek-v4-pro, attains an execution accuracy of 80.42\% on average across three evaluators, compared with 78.06\% for the manual baseline. The automated approach consistently improves performance over the baseline without views by drastically reducing the reasoning load. For instance, the proportion of queries classified as easy increases from 22.78\% in the baseline without views to 90.51\% when using views generated by GPT-5.5, whereas the manual views only raise this proportion to 67.72\%.

Furthermore, manual semantic modeling incurs prohibitive setup and operational costs. The manual authoring process demands 28,689 seconds for the three analysts on a single small database, whereas the automated agents complete the task in 29 to 116 seconds, which represents a reduction in setup time of two to three orders of magnitude. In addition, manual views tend to be wider and overly generalized, leading to an inflation of the schema context. Consequently, the manual baseline consumes over 9.41 million prompt tokens and incurs a monetary cost of \$51.67 under the GPT-5.5 evaluator. In contrast, the views generated by DeepSeek-v4-pro consume only 5.11 million tokens and cost \$30.94, demonstrating a significant reduction in operational expenditure. In enterprise environments characterized by vast and continuously evolving schemas, manual semantic modeling becomes unscalable. The automated View Generation Agent, augmented by the iterative error feedback module, successfully ingests domain knowledge and adapts to schema changes dynamically, thereby providing a highly efficient and scalable alternative to manual configurations.

\subsubsection{Evaluation of Hierarchical Schema Linking Module}

\begin{table}[ht]
\caption{Performance of Hierarchical Schema Linking module on production dataset. $\Delta$ Tokens is the average number of additional tokens RASL consumes per schema-linking query compared with CoeusBI.} 
\begin{center}
\label{tableschemalinking}
\resizebox{\columnwidth}{!}{
\begin{tabular}{lcccc}
\noalign{\hrule height 1.2pt}
Business Line                     & Metric       & RASL                         & CoeusBI & $\Delta$ Tokens (Avg.)    \\ \hline
                                  & Accuracy & {\color[HTML]{0D1239} 88\%}  & \textbf{98\%}    &                        \\ 
\multirow{-2}{*}{BD-Baijiahao}    & Time     & {\color[HTML]{0D1239} 6.17s} & \textbf{4.3s}    & \multirow{-2}{*}{9782} \\ \hline
                                  & Accuracy & {\color[HTML]{0D1239} 92\%}  & \textbf{100\%}   &                        \\ 
\multirow{-2}{*}{BD-Baidu App}    & Time     & {\color[HTML]{0D1239} 7.21s} & \textbf{4.28s}   & \multirow{-2}{*}{4038} \\ \hline
                                  & Accuracy & {\color[HTML]{0D1239} 90\%}  & \textbf{98\%}    &                        \\ 
\multirow{-2}{*}{BD-Search}       & Time     & {\color[HTML]{0D1239} 6.53s} & \textbf{4.3s}    & \multirow{-2}{*}{7535} \\ \hline
                                  & Accuracy & {\color[HTML]{0D1239} 93\%}  & \textbf{100\%}   &                        \\ 
\multirow{-2}{*}{BD-Haokan Video} & Time     & {\color[HTML]{0D1239} 5.73s} & \textbf{4.83s}   & \multirow{-2}{*}{4785} \\ 
\noalign{\hrule height 1.2pt}
\end{tabular}
}
\end{center}
% \vspace{-1.5em}
\end{table}

Because the wide schemas in production datasets make prompt-based schema linking methods~\cite{pourreza2024din, gao2024xiyan, wang2025mac} infeasible due to limits on tokens, we empirically compare the hierarchical schema-linking module with RASL~\cite{eben2025rasl}. 
We evaluate on four Baidu production datasets with wide schemas (BD-Baijiahao, BD-Baidu App, BD-Search, and BD-Haokan Video) using 235 real-world queries to assess the correctness of schema-linking. 
For privacy reasons, we use DeepSeek-V3~\cite{guo2025deepseek} as the rerank model for both methods. 
Table~\ref{tableschemalinking} shows that RASL has larger schema-linking overhead and slower runtime. 
This stems from applying semantic descriptions at the Query Engine Layer instead of at the Data Modeling Layer of CoeusBI, which introduces extra tokens for each schema-linking query. 
Moreover, these descriptions lack awareness of domain knowledge, reducing accuracy in identifying columns rich in BI-specific semantics. 
In contrast, CoeusBI benefits from accurate view generation and a view-first–then-column schema-linking design, enabling the schema-linking module to meet production requirements. 

\subsubsection{Evaluation of Routing Agent and NL2IR2SQL Pipeline}

\begin{table}[htbp]
\begin{center}
\caption{Evaluations on multi-dialect support on SRD production dataset (SQLite).}
\label{exp:SRDDatasetSqlite}
\begin{tabular}{lccc}
\noalign{\hrule height 1.2pt}
\multicolumn{1}{l}{Method}              & \multicolumn{1}{c}{UEX (\%)}      & \multicolumn{1}{c}{PT}              & RT             \\ \hline
\multicolumn{1}{l}{DIN-SQL + DeepSeek-V3} & \multicolumn{1}{c}{31.6}          & \multicolumn{1}{c}{2958765}         & 105905         \\ 
\multicolumn{1}{l}{MAC-SQL + DeepSeek-V3} & \multicolumn{1}{c}{57.9}          & \multicolumn{1}{c}{3694852}         & 95378          \\ 
\multicolumn{1}{l}{SiriusBI-QwenCoder-2504} & \multicolumn{1}{c}{60.8}          & \multicolumn{1}{c}{3014752}         & 19326          \\ 
\multicolumn{1}{l}{XiYanSQL-QwenCoder-2504}             & \multicolumn{1}{c}{49.1}          & \multicolumn{1}{c}{1495377}         & \textbf{10754} \\ 
\multicolumn{1}{l}{CoeusBI + DeepSeek-V3}               & \multicolumn{1}{c}{\textbf{87.1}} & \multicolumn{1}{c}{\textbf{798753}} & 32465          \\ 
\noalign{\hrule height 1.2pt}
\end{tabular}
\end{center}
% \vspace{-1.5em}
\end{table}

To evaluate the impact of multiple dialects on the performance of the methods, we repeat experiments on the SRD production dataset using SQLite as the underlying database. 
DIN-SQL adopts a few-shot strategy with numerous SQLite-specific, golden SQL examples. 
Table~\ref{exp:SRDDatasetSqlite} shows that DIN-SQL leads to a substantial UEX decline on the SRD production dataset under the ClickHouse dialect compared with the SQLite dialect. 
For methods that leverage few-shot prompting to support database-specific dialects, providing more comprehensive exemplars entails additional costs. 
By contrast, the NL2IR2SQL pipeline of CoeusBI, which is agnostic to dialects, delivers consistently strong results under both dialects.
Moreover, CoeusBI explicitly integrates domain knowledge, thereby achieving markedly superior performance on production datasets. 

\subsection{Ablation Study}

\subsubsection{Ablation Studies on View Generation Agent, Hierarchical Schema Linking Module and NL2IR2SQL Pipeline}\label{evaluation:ab1} 

\begin{table}[htbp]
\begin{center}
\caption{Ablation study of VGA-HSL and NL2IR2SQL pipeline on the BIRD-dev dataset. ``VGA-HSL'' denotes the View Generation Agent and Hierarchical Schema Linking module.}
\label{exp:ablation1}
\begin{tabular}{lcccc}
\noalign{\hrule height 1.2pt}
\multicolumn{1}{l}{Method}              & \multicolumn{1}{c}{EX (\%)}       & \multicolumn{1}{c}{Time (s)}      & \multicolumn{1}{c}{PT}               & RT             \\ \hline
\multicolumn{1}{l}{CoeusBI + DeepSeek-V3}               & \multicolumn{1}{c}{\textbf{71.4}} & \multicolumn{1}{c}{8.91}          & \multicolumn{1}{c}{6112623}          & 187110         \\ 
\multicolumn{1}{l}{\quad w/o VGA-HSL}   & \multicolumn{1}{c}{23.7}          & \multicolumn{1}{c}{\textbf{5.38}} & \multicolumn{1}{c}{8571085}          & 162884         \\ 
\multicolumn{1}{l}{\quad w/o NL2IR2SQL} & \multicolumn{1}{c}{62.8}          & \multicolumn{1}{c}{9.43}          & \multicolumn{1}{c}{\textbf{5185149}} & \textbf{67052} \\ 
\noalign{\hrule height 1.2pt}
\end{tabular}
\end{center}
% \vspace{-2em}
\end{table}

Because the Hierarchical Schema Linking module (HSL) follows a view-first–then-column design, we evaluate it together with the View Generation Agent (VGA) in an ablation configuration, denoted VGA–HSL. 
Table~\ref{exp:ablation1} shows that CoeusBI exhibits a substantial advantage in EX compared with CoeusBI w/o VGA–HSL. 
In terms of E2E Time, CoeusBI w/o VGA-HSL attains lower E2E Time because it bypasses schema linking and treats the entire schema as the linking result. 
These findings underscore the importance of combining VGA and HSL for improving the accuracy of the SQL generation.
VGA transforms join-intensive queries into simple single-view queries, while HSL enables accurate schema linking in wide schema settings. 

We also evaluate the effectiveness of the NL2IR2SQL pipeline. 
For CoeusBI w/o NL2IR2SQL, we employ XiYanSQL-QwenCoder-2504 as the SQL generation model for a fair comparison. 
Table~\ref{exp:ablation1} shows that CoeusBI not only achieves higher EX but also reduces E2E Time compared with CoeusBI w/o NL2IR2SQL, indicating that the NL2IR2SQL pipeline, designed for NL2BI, significantly improves the accuracy of SQL generation on public datasets. 
By recasting NL2SQL as a reduced-search-space NL2IR task and then deterministically translating the IR to SQL, the pipeline improves the accuracy of the generated queries. 
In terms of PT, because production deployments incorporate substantial domain knowledge as input, CoeusBI incurs slightly higher PT than CoeusBI w/o NL2IR2SQL. 
For RT, the standardized SQL output enforced by Xiyan-SQL confers a pronounced advantage to CoeusBI w/o NL2IR2SQL. 

\subsubsection{Ablation and Sensitivity Studies}

\begin{table}[htbp]
\begin{center}
\caption{Ablation study of Schema Linking Modes and Hyperparameters on SRD production dataset.}
\label{exp:ablation_linking}
\begin{tabular}{lc}
\noalign{\hrule height 1.2pt}
Configuration                           & UEX (\%)      \\ \hline
CoeusBI + DeepSeek-V3 (Default $K=50$, $\alpha=0.95$) & \textbf{86.5}     \\ 
\quad w/o C2C mode                      & 49.1                       \\ 
\quad w/o V2C mode                      & 79.5                       \\ 
\quad w/o ALL2C mode                    & 83.0                       \\ 
\quad $K=10$                            & 67.8                       \\
\quad $K=30$                            & 81.9                       \\ 
\noalign{\hrule height 1.2pt}
\end{tabular}
\end{center}
% \vspace{-2em}
\end{table}

\paragraph{Crucial Role of Direct Column Mapping} As shown in Table~\ref{exp:ablation_linking}, the ablation study reveals that the Direct Column Mapping mode serves as the foundational pillar for schema linking. Omitting the C2C mode causes the UEX metric to plummet from 86.5\% to 49.1\%, reflecting a severe degradation in alignment capability. In comparison, removing the V2C and ALL2C modes results in moderate performance regressions, decreasing UEX to 79.5\% and 83.0\% respectively. This disparity indicates that while value-based mappings and domain knowledge paraphrases provide essential supplementary signals to capture implicit query semantics, direct column mentions remain the primary anchoring mechanism for resolving schema elements accurately.

\paragraph{Sensitivity to Retrieval Capacity} The evaluation demonstrates a clear positive correlation between the retrieval size hyperparameter $K$ and overall execution accuracy. Restricting the retrieval candidates by reducing $K$ from the default 50 down to 10 significantly deteriorates the UEX to 67.8\%. A moderate retrieval size of $K=30$ recovers a substantial portion of the performance, achieving an 81.9\% UEX, yet it still trails the optimal baseline configuration. This trend suggests that capturing a wider semantic candidate space during the initial vector database retrieval phase is necessary to mitigate early-stage mapping omissions and ensure sufficient context for downstream resolution.

\subsubsection{Ablation Study of MRD Support Module}

\begin{table}[htbp]
\begin{center}
\caption{Ablation study of MRD support module on MRD production dataset. To isolate the effect of multi-round handling strategies, all configurations employ oracle schema-linking results and explicit MRD routing labels, excluding schema-linking errors and routing misclassifications  (QR = Query rewrite, ALI = All inputs). }
\label{exp:ablation3}
\begin{tabular}{lcccc}
\noalign{\hrule height 1.2pt}
\multicolumn{1}{l}{Method}              & \multicolumn{1}{c}{UEX (\%)}       & \multicolumn{1}{c}{Time (s)}      & \multicolumn{1}{c}{PT}              & RT             \\ \hline
\multicolumn{1}{l}{CoeusBI + DeepSeek-V3}               & \multicolumn{1}{c}{\textbf{83.3}} & \multicolumn{1}{c}{9.23} & \multicolumn{1}{c}{727251}          & \textbf{17925} \\ 
\multicolumn{1}{l}{\quad w/o MRD + QR}  & \multicolumn{1}{c}{74.3}          & \multicolumn{1}{c}{9.52}          & \multicolumn{1}{c}{712654}          & 24293          \\ 
\multicolumn{1}{l}{\quad w/o MRD + ALI} & \multicolumn{1}{c}{75.7}          & \multicolumn{1}{c}{\textbf{8.01}} & \multicolumn{1}{c}{\textbf{663493}} & 21907          \\
\noalign{\hrule height 1.2pt}
\end{tabular}
\end{center}
% \vspace{-1.5em}
\end{table}

Here we present an ablation study of the MRD Support Module using the MRD production dataset, with results summarized in Table~\ref{exp:ablation3}. 
In the CoeusBI w/o MRD + Query rewrite setting, MRD queries are handled by rewriting the current query based on previous questions. 
In the CoeusBI w/o MRD + All inputs setting, all historical queries from the dialogue session are included in the prompt for the current query. 
The standard CoeusBI modifies only the previous golden IR according to the current query. 
Specifically, when the MRD support module is removed, the accuracy of generated results drops by 10.8\% (when using Query rewrite) and by 9.13\% (when using All inputs) compared to the standard CoeusBI. 
These results indicate that CoeusBI strongly benefits from the MRD support module for effective MRD processing. 
In terms of E2E Time, CoeusBI is slightly faster than CoeusBI w/o MRD + Query rewrite, but slightly slower than CoeusBI w/o MRD + All inputs. 
In BI scenarios that require accurate support for MRD, the MRD support module is critical.
By modifying the richer IR rather than rewriting the user query, it ensures accurate handling of the MRDs.

\subsubsection{Utility and Usability Study}

We follow the gray-release evaluation used by SiriusBI~\cite{jiang2024siriusbi}. 
On the data platform of Baidu, we randomly assign 50 business analysts to two groups: Group A (25 analysts), who are granted access to CoeusBI for daily tasks, and Group B (25 analysts), who continue to use legacy tools. 
The experiment runs for three weeks, during which we monitor the E2E query workflows of both groups. 
Group A achieves a 55.7\% reduction in average query completion time compared with Group B. 
This improvement is primarily attributable to the E2E architecture and advanced technical design of CoeusBI. 
Participant feedback highlights two key strengths: 
1) 93\% of users report that the Hierarchical Schema Linking module reliably identifies appropriate schema elements, reducing the need to consult domain knowledge documentation to locate relevant columns. 
2) 89\% of users report that the MRD support module enables efficient, human-like partial modifications without requiring the execution of the entire workflow. 
Overall, the gray-release test indicates that CoeusBI improves both productivity and user experience by addressing functional gaps in existing tooling. 

\section{Related Work}

\subsection{Commercial and Semantic BI Systems}
Modern commercial business intelligence products—such as Snowflake Cortex Analyst, Databricks Genie with AtScale, ThoughtSpot, and Google BigQuery with Gemini—increasingly adopt a semantic layer paired with natural language translation. 
These systems define metrics and dimensions using structured abstractions, effectively mapping user intents to analytical queries. 
However, the establishment of this semantic layer in commercial platforms typically requires meticulous manual definition using YAML or LookML configurations, demanding extensive human labor. 
In contrast, CoeusBI introduces novelty by utilizing a View Generation Agent that automatically generates semantic views from database schemas and domain knowledge, eliminating the manual bottleneck of traditional commercial offerings. 

Academic systems such as SiriusBI~\cite{jiang2024siriusbi} target similar environments by building domain-aware association graphs and employing intermediate representations. 
However, SiriusBI treats the intermediate representation merely as auxiliary context and relies on a language model to generate the final SQL, leaving the system susceptible to hallucination. 
CoeusBI differentiates itself by deterministically compiling the intermediate representation into executable queries, ensuring complete predictability across multiple dialects. 

\subsection{Schema Linking Methods}
Schema linking scales poorly as database size grows~\cite{talaei2024chess}. 
Methods such as DBCopilot~\cite{wang2025dbcopilot} formulate linking as path generation over a hierarchical graph, while CRUSH4SQL~\cite{kothyari2023crush4sql} embeds column names to induce candidate schemas. 
The current state-of-the-art method, RASL~\cite{eben2025rasl}, generates per-table descriptions but injects them directly into prompts at query time, incurring substantial token overhead. 
CoeusBI shifts the application of descriptions to the Data Modeling Layer, using views to implicitly resolve schema ambiguity and entirely avoiding runtime token inflation. 

\section{Conclusion}
In this paper, we present CoeusBI, an interactive BI system deployed at Baidu. 
Unlike commercial systems requiring manual configurations, CoeusBI automates the creation of semantic views to convert complex JOIN operations into simple queries via the View Generation Agent. 
It incorporates a Hierarchical Schema Linking module tailored for wide schemas and implements a Routing Agent paired with a deterministic SQL compiler to ensure accurate support for multi-round dialogues and eliminate hallucination risks in the final SQL generation step. 
CoeusBI serves thousands of users within Baidu, and the strong performance on both public datasets and production datasets demonstrates significant practical effectiveness and scalability.

%\clearpage

\balance
\bibliographystyle{ACM-Reference-Format}
\bibliography{main}

\end{document}